\documentclass{aa}
\usepackage{epsfig,color}
\usepackage[dvips]{rotating}
\begin{document}
\newcommand{\ang}{$\rm \AA$}
\newcommand{\tauross}{$\tau_{\mathrm{ross}}$}
\newcommand{\Msun}{M$_{\odot}$}
\newcommand{\Rsun}{R$_{\odot}$}
\newcommand{\be}{\begin{equation}}
\newcommand{\ee}{\end{equation}}
\newcommand{\bee}{\begin{eqnarray}}
\newcommand{\ad}{$\theta_d$}
\newcommand{\vt}{$\xi_t$}
\newcommand{\cc}{$\mathrm{^{12}C/^{13}C}$}
\newcommand{\kn}{$\kappa_{\nu}$}
\newcommand{\lnu}{$l_{\nu}$}
\newcommand{\km}{$\rm{km\,s^{-1}}$}
\newcommand{\ha}{H$_{\alpha}$}
\newcommand{\ea}{et al.}
\newcommand{\ene}{\end{eqnarray}}
\newcommand{\Halpha}{H$_{\alpha}$\,}
\newcommand{\Hbeta}{H$_{\beta}$\,}
\newcommand{\kms}{km~s$^{-1}$}
\newcommand{\teff}{T$_{\mathrm{eff}}$}
\newcommand{\mic}{$\mu$m}
\newcommand{\astronj}[1]{{AJ }{#1}}
\newcommand{\astrastr}[1]{{A\&A }{#1}}
\newcommand{\nat}[1]{{Nature }{#1}}
\newcommand{\ica}[1]{{Icarus }{#1}}
\newcommand{\sci}[1]{{Science }{#1}}
\newcommand{\aass}[1]{{A\&AS }{#1}}
\newcommand{\apj}[1]{{ApJ }{#1}}
\newcommand{\apjs}[1]{{ApJS }{#1}}
\newcommand{\mnras}[1]{{MNRAS }{#1}}
\newcommand{\acas}[1]{{Acta Astron.\ }{#1}}
\newcommand{\ass}[1]{{Ap\&SS }{#1}}
\newcommand{\geo}[1]{{Geophys.\ J.\ }{#1}}
\newcommand{\geoc}[1]{{Geoch.\ Cosmoch.\ Acta }{#1}}
\newcommand{\mop}[1]{{Moon\ Planets }{#1}}
\newcommand{\jcp}[1]{{J.\ Chem.\ Phys.\ }{#1}}
\newcommand{\pasj}[1]{{PASJ }{#1}}
\newcommand{\pasp}[1]{{PASP }{#1}}
\newcommand{\pasa}[1]{{Proc.\ Astron.\ Soc.\ Aust.\ }{#1}}
\newcommand{\baic}[1]{{Bull.\ Astron.\ Inst.\ Czechosl.\ }{#1}}
\newcommand{\bain}[1]{{Bull.\ Astron.\ Inst.\ Neth.\ }{#1}}
\newcommand{\memras}[1]{{Mem.\ R.\ Astron.\ Soc.\ }{#1}}
\newcommand{\dao}[1]{{Pub.\ Dominion Astrophys.\ Obs.\ }{#1}}
\newcommand{\ibvs}[1]{{Inf.\ Bull.\ Variable Stars\ }{#1}}
\newcommand{\ssr}[1]{{Space Sci.\ Rev.\ }{#1}}
\newcommand{\mnassa}[1]{{Mon.\ Notes\ Astron.\ Soc.\ S.\ Afr.\ }{#1}}
\newcommand{\via}[1]{{Vistas in Astron.\ }{#1}}
\newcommand{\anrevaa}[1]{{ARA\&A }{#1}}
\newcommand{\areps}[1]{{Annu.\ Rev.\ Earth Planet.\ Sci.\ }{#1}}
\newcommand{\astrnachr}[1]{{Astron.\ Nachr.\ }{\bf #1}}
\newcommand{\jmolspec}[1]{{J.\ Mol.\ Spectrosc.\ }{#1}}
\newcommand{\apo}[1]{{Appl.\ Optics }{\bf #1}}
\newcommand{\hoogte}[1]{\rule{0pt}{#1}}
\newcommand{\hminff}{H$^-_{\rm{ff}}$}

\def\offinterlineskip{\baselineskip=-1000pt \lineskip=1pt
\lineskiplimit=\maxdimen}
\def\pra{\mathrel{\mathchoice {\vcenter{\offinterlineskip\halign{\hfil$\displaystyle##$\hfil\cr\propto\cr\sim\cr}}}
{\vcenter{\offinterlineskip\halign{\hfil$\textstyle##$\hfil\cr\propto\cr\sim\cr}}}
{\vcenter{\offinterlineskip\halign{\hfil$\scriptstyle##$\hfil\cr\propto\cr\sim\cr}}}
{\vcenter{\offinterlineskip\halign{\hfil$\scriptscriptstyle##$\hfil\cr\propto\cr\sim\cr}}}}}

\def\ga{\mathrel{\mathchoice {\vcenter{\offinterlineskip\halign{\hfil$\displaystyle##$\hfil\cr>\cr\sim\cr}}}
{\vcenter{\offinterlineskip\halign{\hfil$\textstyle##$\hfil\cr>\cr\sim\cr}}}
{\vcenter{\offinterlineskip\halign{\hfil$\scriptstyle##$\hfil\cr
>\cr\sim\cr}}}
{\vcenter{\offinterlineskip\halign{\hfil$\scriptscriptstyle##$\hfil\cr>\cr\sim\cr}}}}}

\def\la{\mathrel{\mathchoice {\vcenter{\offinterlineskip\halign{\hfil$\displaystyle##$\hfil\cr<\cr\sim\cr}}}
{\vcenter{\offinterlineskip\halign{\hfil$\textstyle##$\hfil\cr<\cr\sim\cr}}}
{\vcenter{\offinterlineskip\halign{\hfil$\scriptstyle##$\hfil\cr
<\cr\sim\cr}}}
{\vcenter{\offinterlineskip\halign{\hfil$\scriptscriptstyle##$\hfil\cr><cr\sim\cr}}}}}

\thesaurus{6(08.01.3, 08.06.3, 08.09.2 Alpha Tau, 08.12.1, 13.09.6)}

\title{ISO-SWS calibration and the accurate modelling of cool-star atmospheres:
I. Method\thanks{Based on observations with ISO, an ESA project with instruments
funded by ESA Member States (especially the PI countries France, Germany, the
Netherlands and the United Kingdom) and with the participation of ISAS and NASA.}}
\author{L.\,Decin\inst{1}\thanks{\emph{Scientific researcher of the Fund for
Scientific Research, Flanders}} \and C.\,Waelkens\inst{1} \and
K.\,Eriksson\inst{2} \and B.\ Gustafsson\inst{2} \and B.\ Plez\inst{3} \and
A.J.\ Sauval\inst{4} \and W. Van Assche\inst{5}\thanks{\emph{Research Director
of the Fund for Scientific Research, Flanders}} \and B. Vandenbussche\inst{1}}
\institute{Instituut voor Sterrenkunde, KULeuven, Celestijnenlaan 200B,
    B-3001 Leuven, Belgium \and Astronomiska Observatoriet, Box 515, S-75120
Uppsala, Sweden \and GRAAL - CC72, Universit\'{e} de Montpellier II, F-34095
Montpellier Cedex 5, France \and
Observatoire Royal de Belgique, Avenue Circulaire 3, B-1180 Bruxelles, Belgium
\and Instituut voor Wiskunde, KULeuven, Celestijnenlaan 200B, B-3001 Leuven,
Belgium}

\date{Received 29 November 1999; accepted 3 August 2000}
\offprints{L.\ Decin}
\mail{Leen.Decin@ster.kuleuven.ac.be}
\maketitle
\markboth{L. Decin et al.: ISO-SWS and modelling of cool stars}{L. Decin et al.: ISO-SWS and modelling of cool stars}

\begin{abstract}

A detailed spectroscopic study of the ISO-SWS data of the red giant $\alpha$ Tau
is presented, which enables not only the accurate determination of the stellar
parameters of $\alpha$ Tau, but also serves as a critical review of the ISO-SWS
calibration.

This study is situated in a broader context of an iterative process in which
both accurate observations of stellar templates and cool star atmosphere models
are involved to improve the ISO-SWS calibration process as well as the
theoretical modelling of stellar atmospheres. Therefore a sample of cool stars,
covering the whole A0 -- M8 spectral classification, has been observed in order
to 
disentangle calibration problems and problems in generating the theoretical
models and corresponding synthetic spectrum.

By using stellar parameters found in the literature large discrepancies were
seen between the ISO-SWS data and the generated synthetic spectrum of $\alpha$
Tau. A study of the influence of various stellar parameters on the theoretical
models and synthetic spectra, in conjunction with the Kolmogorov-Smirnov test to
evaluate objectively the goodness-of-fit, enables us to pin down the stellar
parameters with a high accuracy: \teff\ $= 3850 \pm 70$ K, $\log$ g$ = 1.50 \pm
0.15$, M $= 2.3 \pm 0.8$ \Msun, z $= -0.15 \pm 0.20$ dex, \vt\ $= 1.7 \pm 0.3$
\km, \cc\ $= 10 \pm 1$, $\varepsilon$(C) $= 8.35 \pm 0.20$ dex,
$\varepsilon$(N) $= 8.35 \pm 0.25$ dex, $\varepsilon$(O) $= 8.83 \pm 0.15$ dex
and \ad\ $= 20.77 \pm 0.83$ mas. These atmospheric parameters were then compared
with the results provided by other authors using other methods and/or spectra.


\end{abstract}

\vspace{-0.3cm}

\section{Introduction}

The modelling and interpretation of the ISO-SWS (Infrared Space Observatory
- Short Wavelength Spectrometer) data require an accurate calibration of the
spectrometers (Schaeidt et al. 1996). In the SWS spectral region (2.38 -- 45.2
$\mu$m) the primary standard calibration candles are bright, mostly cool,
stars. The better the behaviour of these calibration sources in the infrared is
known, the more accurate the spectrometers can be calibrated. ISO offered the
first opportunity to obtain continuous mid-infrared spectra between 2.38 and
45.2 \mic\ at a spectral resolving power of $\sim 1500$, not polluted by any
molecular 
absorptions of the earth's atmosphere. Therefore our knowledge on the
mid-infrared behaviour of the stellar calibration sources is limited. Refining
the synthetic reference spectra used to calibrate the SWS can only be done by
refining the model atmospheres of the stars.  A full exploitation of the ISO
data can therefore only result from an iterative process in which both accurate
observations and new modelling are involved. In order to obtain a reliable
convergence in such an iterative process, one needs: (1) a sample of bright
stellar objects spread over a large range in spectral type; (2) a thorough
understanding of cool star atmospheres and of the influence of the various
parameters --- e.g.\ \teff, $\log$ g, and chemical composition --- on the
emergent spectra. With this in mind, stars with spectral types
ranging from A0 -- M8 were observed with ISO-SWS. It is important to cover a
broad 
parameter space in order to distinguish calibration problems from
problems related to the model and/or in the generation of the synthetic
spectrum. Concerning point 2 above, it is necessary to perform an
intensive study of the influence of different stellar parameters on synthetic
spectra.  The results of such a study will be presented with the red giant
$\alpha$ Tau as a test case. The results for 15 other stars in our sample will
be presented in forthcoming papers.

This research has been done within the framework of the proposal STARMOD (Title:
Accurate modelling of cool-star atmospheres; P.I.: C. Waelkens; C.I.: M.  Cohen,
L. Decin, Th. de Graauw, L.B.F.M. Waters) and the ZZ\-STARM proposal (Title:
Accurate modelling of cool-star atmospheres; P.I.: L. Decin; C.I.: M. Cohen,
C. Waelkens, Th. de Graauw, L.B.F.M.  Waters). Some calibration data have been
provided by the SWS Instrument Dedicated Team (SIDT) in the framework of a
quick-feedback refining of the model Spectral Energy Distribution (SED) of the
calibration sources used for the SWS calibration.

So far, the analysis of the discrepancies between the ISO-SWS data
and the corresponding synthetic spectra has been restricted to the
wavelength region from 2.38 to 12 \mic, since the lack of
comprehensive molecular and atomic line lists hamper fast progress at
longer wave\-lengths (12 -- 45 \mic). Furthermore the
brightness of the stars drops quickly in this wavelength region so
that the same signal to noise ratio will not be achieved. A third point is that
the SEDs may also be affected by unknown circumstellar contribution.

This paper is organized as follows: in Sect. 2 the observations
are described and the data reduction procedures are discussed. In
Sect. 3 a summary of the literature concerning $\alpha$ Tau is
presented, on the basis of which the starting values for our modelling
stellar parameters are selected. The effect of changing
stellar parameters on the synthetic spectra of K and M giants is
analyzed in Sect. 4. In Sect. 5 the method of analysis is outlined, while in
Sect. 6 the results are discussed. In the last section, 
Sect. 7, the main conclusions are summarized.

\section{Observations and data reduction}\label{obs}

A full scan observation of $\alpha$ Tau has been obtained in the
2.38 -- 45.2 \mic\ wavelength range with the Short-Wavelength
Spectrometer (SWS) (de Graauw et al. 1996) on board ISO. This
has been done by use of the SWS observing mode AOT01 (= a single
up-down scan for each aperture/order combination) with scanner
speed 4, resulting in a resolving power of $\sim$ 1500.  The
observation lasted for 6650 s and was performed during revolution
636.

The data were processed to a calibrated spectrum sam\-pled in all the pixels,
the so-called Auto-Analysis product, using the procedures and calibration files
of the ISO off-line pipeline version 7.0.  For a description of the flux and
wavelength calibration, we refer to Schaeidt et al.\ (1996) and Valentijn et
al.\ (1996). This reduction resulted in an oversampled spectrum for each of the
four AOT-bands, which each consist of 12 detectors.

Remaining instrumental effects, such as fringes in the 12.0 -- 29.5
\mic\ part of the spectrum, were removed using the SWS Interactive
Analysis package (IA) provided by the SIDT. The band-2 data (4.08 -- 12
\mic) are severely affected by detector memory effects. Time and
flux dependent memory effects occur for all four instruments (SWS,
LWS, PHOT, CAM) on board ISO. So far, several attempts have been
made to model the corrections of the transient effects (see e.g.\
Van Malderen et al. 1999), but currently no foolproof method
exists to correct the SWS data for these transient effects. Memory
effects appear to be less severe in the down-scan measurements,
which are obtained after the up-scan data, suggesting a more
stabilized response to the flux level for down-scan data. The SWS
flux calibration of band 2 was corrected for a regime where the
flux history of the detectors is comparable to the second scan in
the observation, the so-called down scan. We therefore used the
down-scan data of our observation as a reference to do a correction of
the flux level of the first scan.

Several procedures were performed on each sub-band separately to
combine the twelve detectors. A sigma-clip\-ping procedure (with $\sigma
=  2.0$) was used to reduce the noise by discarding the datapoints in every
resolution element for which the difference to the mean was $ > 2 \sigma$. After
aligning the twelve different detector signals to their average level
and rebinning to the expected resolution (see Table
\ref{resolution}), the final result was obtained. Although the resolution
changes in within one sub-band, the resolution was kept constant (see
Table \ref{resolution}) and the value was taken to be the lowest value in
Fig.\ 6 by Lorente (1998). The individual
sub-band spectra, when combined into a single spectrum, can show jumps in
flux levels at the band edges. This is due to imperfect flux calibration or
wrong dark-current subtraction for low-flux observations.
Using the overlap regions of the different sub-bands and looking
at other SWS observations, several sub-bands were multiplied by a
small factor (see Table \ref{resolution}) to construct a smooth
spectrum. Due to the problems with memory effects in band 2, the factor of band
2b was determined by use of the template of Cohen et al.\ (1992) and the obtained
synthetic spectra. Note that all shifts are well within the photometric
absolute calibration uncertainties claimed by Schaeidt et al.\ (1996).

\begin{table}
\caption{\label{resolution} Resolution and factors used to shift the sub-bands.}
\vspace{1ex}
\begin{center}
\begin{tabular}{|l|c|c|c|}  \hline
\rule[0mm]{0mm}{5mm} & wavelength & & \\
\rule[-3mm]{0mm}{3mm}{\raisebox{1.5ex}[0pt]{sub-band}} & range [\mic] &
{\raisebox{1.5ex}[0pt]{resolution}} & {\raisebox{1.5ex}[0pt]{factor}}\\
\hline
\rule[0mm]{0mm}{5mm}1a & 2.38 -- 2.60 & 1300 & 1.00 \\
1b & 2.60 -- 3.02 & 1200 & 1.01 \\
1d & 3.02 -- 3.52 & 1500 & 1.00 \\
\rule[-3mm]{0mm}{3mm}1e & 3.52 -- 4.08 & 1000 & 1.00 \\
\hline
\rule[0mm]{0mm}{5mm}2a & 4.08 -- 5.30 & 1200 & 1.00 \\
2b & 5.30 -- 7.00 & 800 & 1.045 \\
\rule[-3mm]{0mm}{3mm}2c & 7.00 -- 12.0 & 800 & 1.00 \\
\hline
\end{tabular}
\end{center}
\end{table}

\section{Literature study: $\alpha$ Tau}\label{liter}

\begin{sidewaystable*}
\caption{\label{literature}Literature study of $\alpha$ Tau, with the effective
temperature \teff\ given in K, the mass M in \Msun, the microturbulent velocity
$\xi_t$ in \km, the angular diameter $\theta_d$ in mas, the luminosity L
in L$_{\odot}$ and the radius R in R$_{\odot}$. Angular diameters deduced from
direct measurements (e.g from interferometry) are written in italic, while
others (e.g.\ from spectrophotometric comparisons) are written upright.}
\vspace{1ex}
\begin{center}
\scriptsize
\begin{tabular}{|c|c|c|c|c|c|c|c|c|c|c|c||r|} \hline
\rule[-3mm]{0mm}{8mm}  \teff & $\log$ g & M & $\xi_t$
&  [Fe/H] & $\varepsilon$(C) & $\varepsilon$(N) & $\varepsilon$(O)
& $^{12}$C/$^{13}$C & $\theta_d$ & L & R & Ref.\\ \hline
\rule[-0mm]{0mm}{5mm}$3910 \pm 200$ & $1.59 \pm 0.30$ & & $2.1 \pm 0.5$ & $-0.34
\pm 0.21$ & & & & & & & & 1.\\
$4140 \pm 100$ & $1.01 \pm 0.46$ & & $2.00 \pm 0.25$ & $-0.33 \pm 0.18$ & $8.27
\pm 0.18$ &
$7.46 \pm 0.28 $ & $8.60 \pm 0.24$ & $9 \pm 1$ & & 200 &  & 2.\\
 & & & & & & & & & 21.154 & & & 3.\\
$3860 \pm 100$ & $1.5 \pm 0.5$ & & $2.2 \pm 0.7 $ & & $8.39 \pm 0.05$ & & & & & & & 4.\\
$3860 \pm 100$ & $1.5 \pm 0.5$ & & $2.2 \pm 0.7 $ & & $8.31 \pm 0.03$ & & & & &
& & 4.\\
$3860 \pm 100$ & $1.5 \pm 0.5$ & & $2.20 \pm 0.14 $ &  & $8.37 \pm 0.14$ & & & & & & & 5.\\
$3850 \pm 100$ & $1.5 \pm 0.3$ & 1.5 & $1.9 \pm 0.3$ & $0.00 \pm 0.20$ & $8.38
\pm 0.08$ & $8.35 \pm 0.06$ & $8.77 \pm 0.08$ & 10 & & & & 6.\\
3831 & 1.41 & 1.5 & 2.0 & & 8.37 & & & $9 \pm 1$ & & & & 7. \\
$3830 \pm 100$ & $1.2 \pm 0.4$ & 1.4 & 2.1 & $-0.14$ & & & & $\sim 12$ &
{\it{24}} & 413 [$463 \pm 59$] & $49 \pm 4$ & 8. \\
3790 & 1.8 & & & & & & & 12 & & & & 9. \\
3898 & 2.0 & & & & & & & & $21.32 \pm 0.58$ & & & 10. \\
3800 & 1.8 & & 1.0 & $-0.17 \pm 0.14$ & & & & & & & & 11. \\
$3970 \pm 70$ & $1.3 \pm 0.35$ & $1.5^{+3.0}_{-1.0}$ & 1.9 & 0.00 & & & & &
{\it{23}} & $414 \pm 75$ &  & 12. \\
3789 & 1.8 & & 1.5 & & & & & $12 \pm 2$ & & & & 13. \\
$3875 \pm 100$ & 1.55 & 1.0 & $2.2 \pm 0.5$ & $+0.16 \pm 0.20$ & $8.73 \pm 0.20$
& $8.61 \pm 0.10$ & $9.02 \pm 0.10$ & $15^{+5}_{-10}$ & & & & 14. \\
$3875 \pm 100$ & 0.55 & 1.0 & $2.2 \pm 0.5$ & $-0.16 \pm 0.20$ & $8.26 \pm 0.20$
& $7.95 \pm 0.10$ & $8.60 \pm 0.10$ & $15^{+5}_{-10}$ & &   & & 14. \\
$3860 \pm 100$ & $1.5 \pm 0.5$ & & $2.5 \pm 1.0$ & & $8.39 \pm 0.05$ & $8.05 \pm
0.25$ & & & & & & 15. \\
$3940 \pm 50$ & $1.25 \pm 0.49$ & & $2.2 \pm 0.2$ & $-0.14 \pm 0.30$ & & & & & &
& & 16.\\
3950 & 1.5 [1.3] & & 1.5 [2.0] & 0.00 & & & $8.97$ [9.03] & & & & & 16. \\
3850 & 0.9 [0.6] & & 1.5 [2.0] & $-0.5$ & & & 8.47 [8.53] & & & & & 16.\\
3750 & 0.5 [0.2] & & 1.5 [2.0] & $-0.5$ & & & 8.27 [8.37] & & & & & 16. \\
 & & & & & & & & & ${\mathit{20.88 \pm 0.10}}$ & & & 17. \\
$3970 \pm 49$ & & & & & & & & & {\it{$20.21 \pm 0.30$}} & & &
18a. \\
$3859 \pm 31$ & & & & & & & & & {\it{21.07}} & & & 18b. \\
 & & & & & & & & & ${\mathit{21.205 \pm 0.21}}$ & & & 19. \\
 & & & & & & & & & ${\mathit{\theta_{\mathrm{UD}} = 19.88 \pm 0.14}}$ & & & 20. \\
 & & & & & & & & & ${\mathit{\theta_{\mathrm{UD}} = 19.80 \pm 0.12}}$ & & & 20. \\
$3976 \pm 485$ & & & & & & & & & & $417 \pm 83$ & $43.0 \pm 4.4$ & 21. \\
$3920 \pm 15$ & & & & & & & & & 20.634 & & & 22. \\
3690 & & & & & & & & & & & & 23. \\
3943 & 1.2 & & & $-0.10$ & & & & & {\it{20.62}} & & & 24. \\
 & & & & $-0.102\pm 0.038$ & & & & & & & & 25. \\
$3790 \pm 200$ & $2.2 \pm 0.3$ & & & $-0.22 \pm 0.3$ & & &
& & & & & 26. \\
 & & & & & & & & & 20.0--35.0 & & 23.0 & 27$^a$. \\
 & & & & & & & & & {\it{16.0--24.0}} & & 45.0--52.0 & 27$^b$. \\
$3947 \pm 41$ & & & & & & & & & ${\mathit{20.44 \pm 0.11}}$ & & & 28. \\
3840 & & & & & & & & & 21.4 & & & 29. \\
3820 & & & & & & & & & $20.3 \pm 1.4$ & & & 30.\\
$3490 \pm 150$ & & & & & & & & & $21.6 \pm 1.9$ & & & 31. \\
4000 & & & & & & & & & $21.4 \pm 1.1$ & & & 32.\\
\rule[-3mm]{0mm}{3mm} $3850\pm 100$ & $1.5 \pm 0.3$ & 1.5 & $1.9 \pm 0.3$ &
$0.00 \pm 0.20$ & $8.40 \pm 0.20$ & $8.20 \pm 0.20$ & $8.78 \pm 0.20$ & $10$ & &
& & 33. \\ \hline
\end{tabular}
\end{center}
\footnotesize{
1. McWilliam 1990;
2. Lambert \& Ries 1981;
3. Blackwell et al.\ 1990;
4. Tsuji 1991;
5. Tsuji 1986;
6. Smith \& Lambert 1985;
7. Harris \& Lambert 1984;
8. Kov\'{a}cs 1983;
9. Lambert et al.\ 1980;
10. Cohen et al.\ 1996b;
11. Fern\'{a}ndez-Villaca\~{n}as et al.\ 1990;
12. van Paradijs \& Meurs 1974;
13. Tomkin \& Lambert 1974;
14. Luck \& Challener 1995;
15. Aoki \& Tsuji 1997;
16. Bonnell \& Bell 1993;
17. Ridgway et al.\ 1982;
18a. Di Benedetto \& Rabbia 1987;
18b. Di Benedetto 1998;
19. Mozurkewich et al.\ 1991;
20. Quirrenbach et al.\ 1993;
21. Volk \& Cohen 1989;
22. Blackwell et al.\ 1991;
23. Linsky \& Ayres 1978;
24. Bell \& Gustafsson 1989;
25. Taylor 1999;
26. Burnashev 1983;
27$^{a,b}$. Fracassini et al.\ 1988;
28. Perrin et al.\ 1998;
29. Engelke 1992;
30. Blackwell \& Shallis 1977;
31. Scargle \& Strecker 1979;
32. Manduca et al.\ 1981;
33. Smith \& Lambert 1990}
\end{sidewaystable*}

\normalsize

$\alpha$ Tau (= HD~29139 = HR~1457 = HIC~21421) is classified as a K5III
giant. A detailed literature study was necessary, on the one hand to extract the
best possible 
set of starting fundamental parameters in order to reduce the number of
calculated spectra and on the other hand to check the consistency
between the stellar parameters deduced from the ISO-SWS data and other
spectra/methods. An exhaustive discussion of published results is presented in
Table \ref{literature} and in Appendix A.

\begin{figure}[h]
\begin{center}
\scalebox{0.35}[0.4]{\rotatebox{90}{\includegraphics{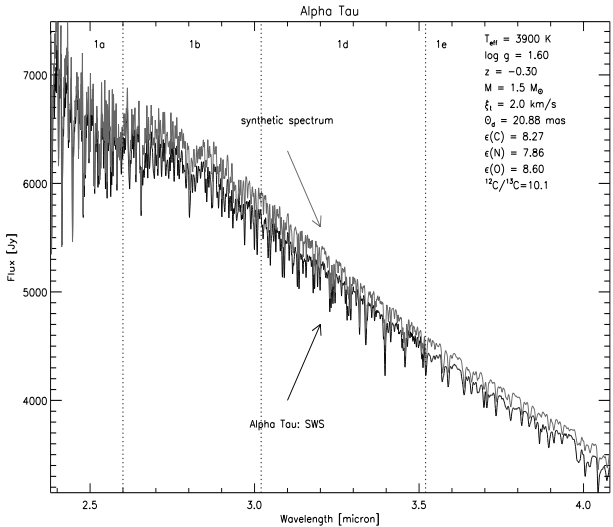}}}
\caption{Comparison between band 1 of the ISO-SWS data of
$\alpha$ Tau (black)
and the synthetic spectrum (grey) with stellar
parameters \teff\ = 3900 K, $\log$ g = 1.60, M = 1.5 \Msun, z =
$-0.30$, \vt\ = 2.0 \km, \cc\ = 10, $\varepsilon$(C) =
8.27, $\varepsilon$(N) = 7.86, $\varepsilon$(O) = 8.60 and \ad\ = 20.88
mas. \label{startvalues}}
\end{center}
\end{figure}

After evaluation of the various sources, the stellar parameters of McWilliam
(1990), with \teff\ = 3900 K, $\log$ g = 1.60, \vt\ = 2.0 \km,
[Fe/H] = $-0.30$, together with $\varepsilon$(C) = 8.27, $\varepsilon$(N) =
7.86, $\varepsilon$(O) = 8.60 (Lambert \& Ries (1981), but with [N/Fe] taken to
be $+0.20$ according to Luck \& Challener (1995), see Appendix A), a mass of 1.5
\Msun\ (van Paradijs \& Meurs 
1974), the limb-darkened angular diameter \ad\ $= 20.88$ mas from Ridgway et al.
(1982) and \cc\ $= 10$ (Smith \& Lambert 1985) were taken as starting values.
In Fig. \ref{startvalues} both  band 1 of the ISO-SWS data of
$\alpha$ Tau and the synthetic spectrum generated with these parameters are
shown. From Fig. \ref{startvalues}, it is already obvious that not only the
angular diameter \ad\ should be lowered and/or \teff\ should be increased, but
also other parameters may need improvement.

\section{Influence of stellar parameters on synthetic spectra}\label{param}

In this section the effect of stellar parameters on the absorption by CO, SiO,
OH and H$_2$O is discussed. We mainly focus on these molecules since they are
the most prominent absorbers for oxygen-rich giants in the wavelength range
considered here (2.38 -- 12 \mic) (see, e.g., Fig. 2 and Fig. 3 in Decin et
al. (1997) and Fig. \ref{atdiv} in this paper).  The goal of this part of the
study is to learn how the synthetic spectrum will change when one of the
parameters is changed within its uncertainties, the only exception being the
stellar mass and the \cc\ ratio due to their small influence. A simple equation
for any chemical compound is not easily obtained, nor is a unique scenario able
to explain all different behaviours, but some patterns do emerge. These results
are then useful for the determination of the origin of the discrepancies seen
between ISO-SWS data and synthetic spectra.

Since the strength of molecular (and atomic) lines is proportional to the ratio
of line to continuous absorption coefficient, $l_{\nu}/\kappa_{\nu}$,
approximations for this ratio are sought for. The approximate ratio is referred
to as ${\cal{R}}$ and equations for some molecules have been deduced according
to the approach of Kj{\ae}gaard et al.\ (1982) (see Appendix B).

The ratio ${\cal{R}}$ has been studied for the surface flux using the following
parameters:\\
\teff\ = 3650 K -- 3850 K -- 4050 K\\
$\log$ g = 1.00 -- 1.50 -- 2.00\\
z = $-0.15$ -- 0.00 -- 0.15\\
M = 1.5 \Msun\ -- 10 \Msun\ -- 15 \Msun\\
\vt\ = 1.0 \km\ -- 2.0 \km \\
\cc\ = 10 -- 4.26\\
$\varepsilon$(C) = 8.24 -- 8.54\\
$\varepsilon$(N) = 8.26 -- 8.56\\
$\varepsilon$(O) = 8.83 -- 9.13\\

All computed spectra are compared with the synthetic spectrum with
parameters:
\teff\ = 3650 K, $\log$ g = 1.00, z = 0.00, M = 1.5 \Msun, \vt\ = 2.0 \km, \cc\
= 10, $\varepsilon$(C) = 8.24, $\varepsilon$(N) = 8.26,
$\varepsilon$(O) = 8.83. If not specified these stellar parameters are used.
In Fig. \ref{overall} the overall results of a change $\Delta$\teff\ = 200 K
(a), $\Delta\log$ g = 0.50 (b), $\Delta$M = 13.5 \Msun\ (c), $\Delta$z = 0.15
(d), $\Delta$\vt\ = 1.0 \km\ (e), $\Delta$\cc\ = 5.84 (f),
$\Delta\varepsilon$(C) = 0.30 (g), 
$\Delta\varepsilon$(N) = 0.30 (h), $\Delta\varepsilon$(O) = 0.30 (i) are
shown. The results mentioned and discussed in this study apply only to models
with similar stellar parameters.

\begin{figure*}[t]
\begin{center}
\scalebox{0.65}[.8]{\rotatebox{90}{\includegraphics{H1877.f2}}}
\vspace{4ex} \caption{Division of the synthetic spectrum with
parameters \teff\ = 3650 K, $\log$ g = 1.00, M = 1.5 \Msun, z =
0.00, \vt\ = 2.0 \km, \cc\ = 10, $\varepsilon$(C) =
8.24, $\varepsilon$(N) = 8.26, $\varepsilon$(O) = 8.83 by the
synthetic spectrum with the same parameters but with {\bf{(a)}} \teff\  =
3850 K, {\bf{(b)}} $\log$ g = 0.50, {\bf{(c)}} M = 15 \Msun, {\bf{(d)}} z =
0.15, {\bf{(e)}} \vt\ = 1 \km, {\bf{(f)}} \cc\ = 4.26, {\bf{(g)}}
$\varepsilon$(C) = 8.54, {\bf{(h)}} $\varepsilon$(N) = 8.56 and {\bf{(i)}}
$\varepsilon$(O) = 9.13. Prominent 
molecular bands are specified in the plot. Plot {\bf{(a)}} has been divided by
the factor $3650/3850$ to compensate for the higher flux of the spectrum with
an effective temperature of 3850 K.\label{overall}}
\end{center}
\end{figure*}

\subsection{Models}\label{models}

The models and corresponding synthetic spectra have been computed
by using the MARCS-code (Gustafsson et al. 1975). Since 1975, this code
has undergone some modifications, the most important ones
being the replacement of the  Opacity Distribution Function (ODF)
technique by the Opacity Sampling (OS) technique, the
possibility to use a spherically symmetric geometry for extended
objects and major improvements
of the line and continuous opacities (Plez et al. 1992).

The common assumptions of spherical or plane-parallel
stratification in homogeneous stationary layers, hydro\-stat\-ic
equilibrium and LTE were made. Energy conservation was required
for radiative and convective flux, where the energy transport due
to convection was treated through a local mixing-length theory
(Henyey et al. 1965). The mix\-ing-length $l$ was chosen as 1.5 $H_p$,
with $H_p$ the pressure scale height, which is a reasonable quantity
to simulate the temperature structure beneath the photosphere
(Nordlund \& Dravins 1990). Turbulent pressure was neglected.
The reliability of these assumptions is discussed by Plez et al.\ (1992).

The synthetic spectra were generated using the TurboSpectrum
program described by Plez et al.\ (1993), and further updated. The program
treats the chemical equilibrium for hundreds of molecules with a
consistent set of partition functions and dissociation energies.
Solar abundances from Anders \& Grevesse (1989) have been
assumed, except for the iron abundance, $\varepsilon$(Fe)$ = 7.51$,
which is in better agreement with the meteoritic value.

The continuous opacity sources considered here are H$^-$, H, Fe,
(H+H), H$_2^+$, H$_2^-$, He~I, He~I$_{\mathrm{ff}}$, He$^-$, C~I, C~II,
C~I$_{\mathrm{ff}}$, C~II$_{\mathrm{ff}}$, C$^-$, N~I, N~II, N$^-$, O~I, O~II,
O$^-$, CO$^-$, H$_2$O$^-$, Mg~I, Mg~II, Al~I, Al~II, Si~I, Si~II,
Ca~I, Ca~II, H$_2$(pr), He(pr), e$^-_{\mathrm{sc}}$, H$^-_{\mathrm{sc}}$, H$_{2
\,\mathrm{sc}}$, where `$pr$'  
stands for `pressure induced' and `sc' for `scattering'.

For the line opacity in the SWS range (2.38 -- 12 \mic) a database of
infrared lines including atoms and molecules has been prepared.
For the atomic lines the data listed by Hirata \& Horaguchi
(1995) have been used, for CO those by Goorvitch  (1994), for SiO
those by Langhoff \& Bauschlicher (1993), for H$_2$O those by J{\o}rgensen
(1994) and Ames (Partridge \& Schwenke 1997), OH lines by Sau\-val
(Melen et al. 1995), Schwenke (1997) and Goldman et al. (1998), NH lines by
Sauval 
(Grevesse et al. 1990; Geller et al. 1991) and CH lines by
Sauval (Melen et al. 1989; Grevesse et al. 1991) and CN lines by Plez
(private communication). The dissociation energy for CN was taken to be 7.76 eV.
An exhaustive discussion on the accuracy
and completeness of infrared spectroscopic line lists can be found in Decin
(2000). From this study, a preference emerged for the H$_2$O line list
by Ames and the OH line list by Goldman. To remove the small difference
between the originally `vacuum' wavelengths for
the ISO observations and `air' wavelengths in spectroscopic
linelists above 200 nm, Edlen's formula  (Edlen 1966) was
extended to the infrared. 

\subsection{Effect of changing the effective temperature}

When the temperature increases, fewer molecules are for\-med
resulting in a smaller line absorption coefficient. At 2.3 \mic,
\kn\ is primarily due to H$^-$ free-free absorption, thus its
value depends directly on the electron pressure $P_{\mathrm{e}}$. Increasing the
temperature causes more ionization events and thus more free electrons,
but at the same time the H$^-$ ion itself is less easily formed.
Fig. \ref{part}a. shows that, for the models under consideration,
this latter effect is the most important one and so \kn\ decreases
when the effective temperature is increased from 3650 K to 3850 K.
Moving inwards from the outer photosphere, the line absorption
coefficient l$_{\nu}$ of all molecules (not H$^-$) reaches its
maximum at the location where the effect of an increasing temperature
overtakes the effect of a higher density. The partial pressure p of H$^-$, being
proportional to $p$(H~I)*$P_{\mathrm{e}}$,
keeps on rising because of the increase of the number of free electrons
at higher temperature. When the line-to-continuum contrast has to
be known, one has to consider the partial pressure of the molecule
divided by ($p$(H~I)*$P_{\mathrm{e}}$) (see Fig. \ref{overall} and
Fig. \ref{part}b).
The simple statement that there are weaker absorption bands for
higher temperatures does not always hold: the approximate ratio
${\cal{R}}$ of CO, SiO, OH and H$_2$O decreases, but for CN
the lower continuous opacity compensates for the lower
line opacity (see Fig. \ref{part}b). Furthermore, the relative population of
rotational
levels strongly depends on T (the maximum population occurs at a
J$_{\mathrm{max}}$-value and a lower T corresponds to a lower
J$_{\mathrm{max}}$-value). Therefore a given line (J) intensity
could be increased or decreased according to its J-value only.

\begin{figure}[h]
\scalebox{0.35}[0.35]{\rotatebox{90}{\includegraphics{H1877.f3}}}
\caption{Comparison between the partial pressure of several
molecules for the models with \teff\ = 3650 K (full line) and \teff\ = 3850 K
(dashed line) and with $\log$ g = 1.00, M = 1.5 \Msun, z =
0.00, \vt\ = 2.0 \km, \cc\ = 10, $\varepsilon$(C) =
8.24, $\varepsilon$(N) = 8.26, $\varepsilon$(O) = 8.83. In plot ({\bf{a}}) the partial
pressure of the molecules is given, while in plot ({\bf{b}})
the partial pressure of the molecules is divided by
$p$(H~I)*$P_{\mathrm{e}}$.\label{part}} 
\end{figure}

\subsection{Effect of changing the gravity and the mass}

Although g and M are closely related, the effect of changing the mass is smaller
than that of changing the gravity, because the first one only affects the
extension and the latter one also changes the pressure structure of the
atmosphere. This is clearly visible in Fig. \ref{overall}. The assumption of
hydrostatic equilibrium links the gas pressure to the surface gravity. In a
cool star, the gas and electron pressure ($P_{\mathrm{g}}$ and $P_{\mathrm{e}}$)
can be written as:
\begin{eqnarray}
P_{\mathrm{g}} & \approx & constant \cdot g^p \ \ {\footnotesize{(0.59 \le p \le
1.05)}} \\ 
P_{\mathrm{e}} & \approx & constant \cdot g^p \ \ {\footnotesize{(0.20 \le p \le
1.03)}}.\label{Pe}
\end{eqnarray}
Due to the higher extension of the atmospheres considered here, this value of
$p$ has a larger range compared to the value discussed by Gray (1992).

When the gravity increases, the electron pressure increases rapidly
(Eq. \ref{Pe}),  resulting
in a higher H$^-$ free-free absorption. Other effects of increasing the gravity
are higher number densities supporting the molecular formation and the fact that
molecules can be formed till higher temperatures, both increasing \lnu. The
final result is that the approximate absorption coefficient ${\cal{R}}$ can
either increase or decrease depending on the relative change of l$_{\nu}$ and
\kn.

Using the approximate relations according to Kj{\ae}rgaard \ea\ (1982) (see
Appendix B), the conclusion is reached that an increase of gravity
leads to a decrease of the strength of the CO, NH and CN lines and to an
increase of the strength of the H$_2$O lines, with no dependence for the OH and
SiO lines. The increasing gravity does lead to a strengthening of the H$_2$O
lines and also of the OH lines while it diminishes the
${l_{\nu}}/{\kappa_{\nu}}$ ratio of CO, NH, SiO and CN
(Fig. \ref{overall}). From the approximate relations for ${\cal{R}}$ no
quantitative conclusion on the parameter dependence can thus be made.

An increase in mass yields the inverse (but smaller) effect on the partial
pressure (and so on the line absorption coefficient \lnu) as an increase in
gravity.

\subsection{Effect of changing the metallicity}

For the computations with a higher metallicity, [atom/H] has been increased from
0.00 dex to 0.15 dex for all atoms except for H, He, C, N and O. Because an
increase in metal abundance leads to a decrease in gas pressure at each
optical depth (as \kn\ increases), the partial pressures of the molecules CO,
OH, N$_2$, H$_2$O and CN consequently decrease, while p(SiO) and p(TiO) are
somewhat higher because of the
higher abundance of Si and Ti respectively. An increase in the overall metal
abundance increases the amount of electrons ($P_{\mathrm{e}}$) and as a consequence also
the value of $\kappa_{\nu}$ formed
by H$^-$. This increase of \kn\ dominates the changes in the spectrum:
all molecular features become fainter at increased metallicities
(with C, N and O kept constant) (Fig. \ref{overall}).
This can also be seen in the equations for ${\cal{R}}$ (see appendix B) for the
extreme case in which all carbon is present in the form of CO with
T$_{{\mathrm{e\rm{ff}}}} \le 4500$ K. The strong metallicity dependence of
H$_2$O in the approximate relation for ${\cal{R}}$ may be well visible,
though this increase (with decreasing z) is small for weak H$_2$O bands.

\subsection{Effect of changing the microturbulent velocity}

The impact of changing the  microturbulent velocity
is most important for the CO lines with a large equivalent width.
When a line is saturated, increasing \vt\ widens the wavelength
range covered by the absorption and reduces the saturation, thus
increasing the total absorption. In the line center of an unsaturated line,
 a smaller microturbulence corresponds to a
higher absorption coefficient at that frequency (see Fig. \ref{overall}).

Decreasing the microturbulent velocity from 2 \km\ to 1 \km\
diminishes the CO first overtone absorption by maximum 10\% and the CO
fundamental by up to 6\,\%, the SiO first overtone by up to 4\,\%, and the SiO
fundamental by up to 3\,\%, the fundamental band of OH by up to 3\,\% and
the ro-vibrational stretching modes ($\nu_1$ and $\nu_3$) and
bending mode ($\nu_2$) of H$_2$O by up to 0.5\,\%.

\subsection{Effect of changing the isotopic ratio \cc}

Red giant atmospheres display altered C, N, and O abundances
and isotopic ratios (in particular \cc), due to dredge-up
episodes.
Suppose that the \cc\ ratio is changed from 10 (case I) to 4.26 (case II). If a
$^{12}$CO line is saturated in case I, its equivalent width will decrease very
few from case
I to case II. If we assume the $^{13}$CO line to be very weak in case I, its
equivalent width will increase very much from case I to case II by a factor
nearly equal to the new/old isotopic ratio. Therefore it is understandable that
a  decrease in \cc\ will consequently cause ${\cal{R}}$(CO) to increase,
although this increase may be very small if the CO bands are weak
(Fig. \ref{overall}).

\subsection{Effect of changing $\varepsilon$(C), $\varepsilon$(N) or
$\varepsilon$(O)}

When the ratio C/O increases (but remains $<$ 1), the structure of
the atmosphere is affected
through changes in line blanketing. More CO will be formed and
less oxygen will be left (after CO formation) to form oxygen-based
molecules. Hence, as illustrated in Fig. \ref{overall}, the total
absorption of CO will increase, while it will decrease for SiO,
OH and H$_2$O. Increasing the oxygen abundance gives roughly
the reverse effect, but the total CO absorption hardly changes.
The largest effect of an increasing nitrogen abundance is an
increase of the NH and CN line strengths.

\section{Method of analysis}

\begin{figure}[h]
\begin{center}
\scalebox{0.5}[0.5]{\rotatebox{90}{\includegraphics{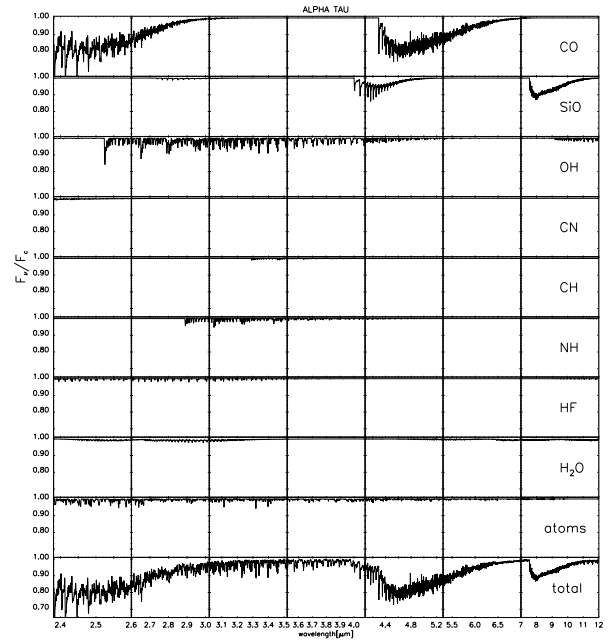}}}
\vspace{2ex}\caption{Relative importance of atoms and molecules in a model with
\teff\ = 3850
K, $\log$ g = 1.00, M = 1.5 \Msun, z =
0.00, \vt\ = 2.0 \km, \cc\ = 10, $\varepsilon$(C) =
8.24, $\varepsilon$(N) = 8.35, $\varepsilon$(O) = 8.83. \label{atdiv}}
\end{center}
\end{figure}

Fig. \ref{startvalues} reveals several discrepancies. As outlined in Sect. 1,
the origin of these discrepancies can be either a calibration problem or a
problem related to the generation of the synthetic spectrum. Precisely because
this research involves both theoretical developments on the model spectra and
calibration improvements of the spectral reduction method, one has to be
extremely careful not to confuse technical detector problems with astrophysical
issues. Therefore, several points have to be taken into account when
scrutinizing these discrepancies.

First of all, one has to take into account that the calibration
process is a delicate one. Due to the high flux level of the observations,
dark-current subtraction does not influence the final spectrum much,
but memory effects affect the reliability of band-2 observations.
Therefore, mainly the band-1 data were used for the determination of the
stellar parameters. The final result is however very sensitive to
the Relative Spectral Response Function (RSRF) used. The RSRF was
intensively measured in the lab before launch by use of a
cryogenic blackbody source with temperatures between 30 and 300K.
After the launch of ISO these intermediate RSRFs were corrected
for broad-band residues detected by comparing reference spectral energy
distributions (SEDs) --- synthetic spectra (Kurucz 1993; van der Bliek et al.
1998) and composites of M. Cohen (Cohen et
al. 1992, 1996a; Witteborn et al. 1999) ---
with SWS observations processed with these intermediate RSRFs
using stellar candles of different IR spectral signature. By using
a cubic spline smoothing algorithm and sigmaclipping the risk was
minimized of introducing false spectral features and one also
avoids possible bias by spectral features of one type of source (Vandenbussche,
1999). The absolute flux level is determined by use of photometric data
and composite SEDs of various sources. The flux calibration
accuracy is estimated to be 5 -- 7\,\% in band 1, 7 -- 12\,\% in band 2a,
7 -- 15\,\% in band 2b and 11 -- 25\,\% in band 2c, the worst accuracy being
located at the edges. The spectral cover from A0 -- M7 was highly necessary,
since 
the different structures of the various selected stars make it possible to
decide upon the quality of the used RSRFs. To test our findings, 
high-resolution observations of two stars were also studied in detail. For a
discussion about the quality of the RSRFs, we refer to the forthcoming papers
which will describe the other 15 stars in our sample.

In order to elucidate problems with the theoretical atmospheric structure, it is
important to know the relative importance of the different molecules (see Fig.
\ref{atdiv}) and the influence of the stellar parameters on the
total absorption by these molecules (see Sect. \ref{param}).  Fig.
\ref{atdiv} shows clearly that especially the molecules CO, OH and
SiO are helpful in determining the atmospheric parameters. In spite of the
moderate resolution of ISO-SWS, we will demonstrate that one can pin down the
stellar parameters of cool giants very accurately from these data. This is due
to the large wavelength range of ISO-SWS, where different important molecules
determine the spectral signature. Since these different spectral features do
each react in another way on a change of one of the several heterogeneous
stellar parameters, it is possible to improve on the stellar
parameters. Using a metallicity of z $= -0.30$
yields, e.g., a synthetic spectrum which has too low a flux density in the
wavelength range from 2.6 -- 2.8 \mic, with the line contrast
${l_{\nu}}/{\kappa_{\nu}}$ being too strong. Fig. \ref{overall} turned
out to be extremely useful to ascertain the gravity: the inverse
influence of the gravity on the OH lines compared to the CO and
SiO lines, together with the steep slope in band 1B and band 1D
make it possible to pin down the logarithm of the gravity within
0.15 dex! Because the microturbulent velocity \vt\ does not only act
on the CO lines, but also on the strongest SiO and OH lines,
one may disentangle a wrong microturbulence and a wrong carbon
abundance.

In order to achieve the highest possible agreement be\-tween the data and the
different synthetic spectra, a statistical approach is proposed.
A choice was made for the Kolmogorov-Smirnov test.
This well-developed goodness-of-fit criterion is applicable for a
broad range of comparisons between two samples, where the kind of
differences which occur between the samples can be very diverse.
An elaborate discussion about the Kolmogorov-Smirnov statistics
can be found in Pratt \& Gibbons (1981) and H\'{a}jek (1969). In
this goodness-of-fit test a parameter, $V_k$, is defined in the
present case as the ratio of the ISO-SWS flux to the synthetic
flux at frequency point $\nu_k$. These parameters  $V_k$ are assumed to be
independent 
random variables with $E\{V_k\} = 1$ and $E\{V_k^2\} = \sigma^2$, with a
constant unspecified variance $\sigma^2$. The random variables
\begin{eqnarray}
Y_i \equiv
\frac{\sum\limits_{k=1}^{i}V_k}{\sum\limits_{k=1}^{n}V_k} & &
(i=1, \ldots, n-1)
\end{eqnarray}
(with $n$ the total number of frequency points) are then
asymptotically behaving like $i/n$ and hence are asymptotically
equidistributed on the interval $[0,1]$.  These variables $Y_i$ will be formally
compared with the quantiles of the uniform distribution $F(x)=x$, $0 \leq x \leq
1$, using a modification of the Kolmogorov-Smirnov test.
Taking the unspecified distribution function of the sample $Y_1, \ldots, Y_n$
as $G$, the null hypothesis is $F=G$. The Kolmogorov-Smirnov test
is based on the maximum difference. Specifically, the test
statistics are given by
\begin{equation}
D_{nn} = \max_{t}|G_n(t)-F_n(t)|.
\end{equation}
One rejects $F=G$ if $D_{nn}$ is `too
large' in comparison to an appropriate critical value. 

Practically, our modification involves the calculation of the value $\beta$,
defined as 
\begin{equation}
\beta \equiv \sqrt{n} \sup\limits_{1 \le i \le
n-1}\left|Y_i-\frac{i}{n}\right|.
\end{equation}
Firstly, we will compute the asymptotical distribution function of $\beta$. By
defining 
\begin{equation}
X_i = \frac{V_i - 1}{\sigma},
\end{equation}
one has that $E\{X_i\} = 0$ and $E\{X_i^2\} = 1$. So,
\begin{eqnarray}
\lefteqn{\beta \equiv \sqrt{n} \sup\limits_{1 \le i \le
n-1}\left|Y_i-\frac{i}{n}\right|  = \sqrt{n} \sup\limits_{1 \le i \le
n-1}\left|\frac{\sum\limits_{k=1}^{i} V_k}{\sum\limits_{k=1}^{n} V_k} - \frac{i}{n}\right|}
\\ 
& & = \sqrt{n} \sup\limits_{1 \le i \le
n-1}\left|\frac{ \displaystyle \frac{\sigma}{n} \sum\limits_{k=1}^{i} X_k -
\frac{i}{n}\left(\frac{\sigma}{n} \sum\limits_{k=1}^{n} X_k \right)}
{ \displaystyle 1 + \frac{\sigma}{n} \sum\limits_{k=1}^{n} X_k}\right|.
\label{eqlong}
\end{eqnarray}
By defining $k \equiv [nt]$, where $[nt]$ is the greatest integer less than or
equal to $nt$, Eq. (\ref{eqlong}) becomes 
\begin{equation}
\beta = \sup\limits_{0 \le t \le 1} \left| \frac{\displaystyle
\frac{\sigma}{\sqrt{n}} \sum\limits_{k = 1}^{[nt]} X_k - 
\frac{[nt]}{n} \left( \frac{\sigma}{\sqrt{n}} \sum\limits_{k=1}^{n} X_k \right)
}{\displaystyle 1 + \frac{\sigma}{n} \sum\limits_{k=1}^{n} X_k}\right|.
\end{equation}
Using the strong law of large numbers
\begin{equation}
 \lim_{n \to \infty} \frac1n \sum_{k=1}^n X_k = 0,
\qquad \textrm{almost surely} 
\end{equation}
and Donsker's theorem (Billingsley, 1968, p.~137)
\begin{equation}  
\frac{1}{\sqrt{n}} \sum_{k=1}^{[nt]} X_k 
   \stackrel{\mathcal{D}}{\to} W(t), 
\end{equation}
where $W(t)$, $0 \leq t \leq 1$, is
standard Brownian motion, i.e., $W(t)$ is a Gaussian process with $W(0)=0$,
independent increments, and $W(t)-W(s)$ has a normal distribution with mean zero
and variance $|t-s|$, one sees that $\beta$ converges in distribution to
\begin{equation}  
\beta \stackrel{\mathcal{D}}{\to} \sigma \sup_{0 \leq t \leq 1} |W(t)-tW(1)|. 
\end{equation}
Here $W_0(t) \equiv W(t)-tW(1)$, $0 \leq t \leq 1$, is known as a Brownian
Bridge. The definition of $W_0$ implies that $W_0$ is a Gaussian process with
$E\{W_0(t)\} = 0$ and \\$E\{W_0(s) W_0(t)\} = s(1-t)$ if $s \le t$. As a
consequence
\begin{equation}  
\beta^* \equiv \frac{\beta}{\sigma}   \stackrel{\mathcal{D}}{\to} \sup_{0 \leq t
\leq 1} |W_0(t)|, 
\end{equation}
which implies that the asymptotic distribution function of $\beta^*$ is given by
the Kolmogorov-Smirnov distribution
\begin{equation}    
P(\beta^* \geq \lambda) = 2 \sum_{i=1}^\infty
   (-1)^{i+1} \exp\{-2i^2\lambda^2\} 
\end{equation}
(Billingsley, 1968, p.~105).
For a test with significance level $\alpha$ one rejects the
null hypothesis when $\beta^* > \lambda_\alpha$, where
\[   2 \sum_{i=1}^\infty
   (-1)^{i+1} \exp\{-2i^2\lambda_\alpha^2\} = \alpha .\]
For $\alpha=0.05$ one has $\lambda_\alpha=1.36$ and for 
$\alpha=0.01$ one has $\lambda_\alpha=1.63$.
So, a good approximation to
the level-$\alpha$ Kolmogorov-Smirnov test is to reject a fit
if the empirical distribution function exits from the
bounds
\begin{equation}
y = x \pm \lambda_{\alpha}(n-1)^{-1/2},\ 0<x<1
\end{equation}
(see, e.g., Fig. 10.2 in Brockwell \& Davis (1991)).

The lower the $\beta$-value, the better the accordance be\-tween
the observed data and synthetic spectrum. It is clear that $\beta = \sigma
\beta^*$ gives us not only information about the correspondence between the
observed and synthetic spectra, but also about $\sigma$. Since $V_k$ is the
ratio of the ISO-SWS flux ($\cal{F}_{\mathrm{obs}}$) to the synthetic flux
($\cal{F}_{\mathrm{syn}}$), it is obvious that 
\begin{equation}
E( {\cal{F}}_{\mathrm{obs}} - {\cal{F}}_{\mathrm{syn}} )^2 =
{\cal{F}}_{\mathrm{syn}}^2 \sigma^2,
\end{equation}
and thus
\begin{equation}
 \sigma^2 = \frac{E( {\cal{F}}_{\mathrm{obs}} - {\cal{F}}_{\mathrm{syn}} )^2}
 {{\cal{F}}_{\mathrm{syn}}^2}.  \label{sigma2}
\end{equation} 
The unknown variance $\sigma^2$ is thus an indication on the variance of the
observed data.

The Kolmogorov-Smirnov
test {\it{globally}} checks the goodness of fit of the observed
and synthetic spectra. An advantage of this test is that one very
discrepant frequency point (e.g.\ due to a wrong oscillator
strength) only mildly influences the final result. To avoid too
high an influence of the less reliable calibration at the band
edges, the parameter $\beta$ was computed for each sub-band
separately. In each sub-band, several sub-intervals were
considered to the main absorber in the relevant wavelength range,
and the parameter $\beta$ was computed for each of this
sub-intervals. The resulting ($Y-F$)-figures turned out to be very
useful to reveal systematic problems (see, e.g., Fig.
\ref{statistic}).

\begin{figure*}[t]
\begin{center}
\scalebox{0.9}[0.8]{{\includegraphics{H1877.f5}}}
\caption{Fig. {\bf{(a-1)}} shows the comparison between band 1 of the
ISO-SWS data of $\alpha$ Tau (black) and the synthetic spectrum
(red) with stellar parameters \teff\ = 3900 K, $\log$ g = 1.60, M =
1.5 \Msun, z = 0.00, \vt\ = 2.0 \km, \cc\ = 10,
$\varepsilon$(C) = 8.24, $\varepsilon$(N) = 8.35, $\varepsilon$(O)
= 8.83 and \ad\ = 20.67; in Fig. {\bf{(a-2)}} the synthetic spectrum has
as stellar parameters \teff\ = 3850 K, $\log$ g = 1.50, M = 1.5
\Msun, z = -0.15, \vt\ = 2.0 \km, \cc\ = 10, $\varepsilon$(C) =
8.24, $\varepsilon$(N) = 8.35, $\varepsilon$(O) = 8.83 and \ad\ =
20.77 mas. In Fig. {\bf{(b-1)}} and {\bf{(b-2)}} the ratio of the ISO-SWS flux
to the synthetic flux of respectively {\bf{(a-1)}} and {\bf{(a-2)}} is plotted.
The difference between the partial sum $Y_i$ and the straight line
$i/n$ is plotted against the number i of the frequency point in
Fig. {\bf{(c-1)}} and Fig. {\bf{(c-2)}}. The parameter $\beta$ of the
different sub-bands is indicated in Fig. {\bf{(b-1)}} and Fig. {\bf{(b-2)}},
whereas the parameter $\beta$ of the entire band 1 is indicated in Fig.
{\bf{(c-1)}} and Fig. {\bf{(c-2)}}. \label{statistic}}
\end{center}
\end{figure*}

\begin{figure}[h]
\begin{center}
\scalebox{0.3}[0.3]{\rotatebox{90}{{\includegraphics{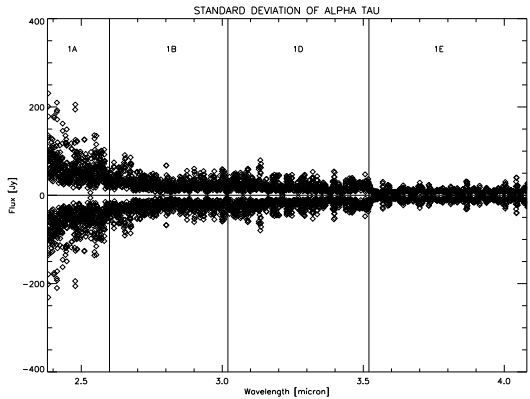}}}}
\caption{Standard deviation of the different bins, obtained when rebinning the
oversampled spectrum of $\alpha$ Tau to the expected resolution.\label{stdevat}}
\end{center}
\end{figure}

Very low $\beta$-values ($< 0.09$) were achieved, proving a good relation
between observed and synthetic spectra.  The objective
$\beta$-parameters, which are - in se - deviation estimating
parameters, were then further used to improve the stellar
parameters. A sensible rule of thumb would then be to select the
synthetic spectrum with the lowest $\beta$-values, among those
that are consistent with the null hypothesis. In addition, it is a
useful sensitivity analysis to compare various spectra with
acceptable $\beta$-values in key characteristics. The high flux
accuracy enables us to distinguish between $\beta = 0.15$ and
$\beta = 0.03$ (see Fig. \ref{statistic}), although both values
are in the same region of accepting the null hypothesis. From the
sensitivity analysis, it seemed to be useful to determine a
maximum acceptable $\beta$-value (i.e.\ a minimal deviation) as an
objective criterion. A synthetic spectrum is accepted to represent
the true stellar parameters of $\alpha$ Tau (and the other stars) when the
$\beta$-values are lower than the values given in column 2 of
Table \ref{beta}. These empirical $\beta_{\mathrm{max}}$-values
give, as already mentioned (Eq. (\ref{sigma2})), an indication on the
uncertainty on the observational spectrum 
determined by e.g.\ the pointing accuracy, the pointing jitter of
ISO, uncertainties on the RSRFs, problematic dark-current
subtraction, ... In Fig. \ref{stdevat} the standard deviation,
obtained when rebinning the oversampled spectrum of $\alpha$ Tau
to the expected resolution, of each bin is shown. Obviously, the
ISO-SWS spectrum is less accurate in band 1A than in the other
bands. This is reflected in the higher maximum acceptable
$\beta$-value in Table \ref{beta}. In column 3 of Table \ref{beta}, the
final $\beta$-values for $\alpha$ Tau are given.

\begin{table}
\caption{\label{beta}$\beta$-value for the different sub-bands and
molecules in that sub-band. In the second column the maximum
acceptable value is given; in the third column the correspondent
value for $\alpha$ Tau.} \vspace{1ex}
\begin{center}
\begin{tabular}{|l|c|c|}  \hline
\rule[0mm]{0mm}{5mm} sub-band/ & maximum acceptable & $\beta$-value \\
\rule[-3mm]{0mm}{3mm} molecule & $\beta$-value & for $\alpha$ Tau \\
\hline \hline
\rule[0mm]{0mm}{5mm}1a & 0.06 & 0.040 \\
1b & 0.05 & 0.034 \\
1d & 0.04 & 0.036 \\
\rule[-3mm]{0mm}{3mm}1e & 0.04 & 0.027 \\
\hline
\rule[0mm]{0mm}{5mm}CO (1a) & 0.06 & 0.040 \\
CO (1b) & 0.05 & 0.039 \\
OH (1b) & 0.04 & 0.024 \\
OH (1d) & 0.04 & 0.036 \\
OH (1e) & 0.03 & 0.025 \\
\rule[-3mm]{0mm}{3mm}SiO (1e) & 0.03 & 0.022 \\
\hline
\end{tabular}
\end{center}
\end{table}

Due to the smaller weight which are given automatically to small features,
the traditional comparison between observed and synthetic spectra by eye-ball 
fitting is still necessary as a complement to this Kolmogorov-Smirnov
method in order to reveal systematic errors in those features. The
final error bars on the atmospheric parameters are then estimated
from 1. the intrinsic uncertainty on the synthetic spectrum (i.e.\
the possibility to distinguish different synthetic spectra at a
specific resolution i.e.\ there should be a relevant difference in
$\beta$-values) which is thus dependent on both the resolving
power of the observation and the specific values of the
fundamental parameters, 2. the uncertainty on the ISO-SWS spectrum
which is directly related to the quality of the ISO-SWS
observation , 3. the value of the $\beta$-parameters in the
Kolmogorov-Smirnov test and 4. the still remaining discrepancies
between observed and synthetic spectrum. However, no exact formula can be given
to compute the error bars due to the dependence of the several parameters.
When a quantity $Q$ depends on other independent parameters, e.g.\ $a$ and $b$,
the uncertainty $s$ is estimated from the expressions

\begin{equation}
s_{<Q>} \simeq \sqrt{\left(\frac{\partial Q}{\partial a}\right)^2
s^2_{<a>} + \left(\frac{\partial Q}{\partial b}\right)^2
s^2_{<b>}}
\end{equation}
or, equivalently
\begin{equation}
\frac{s_{<Q>}}{Q} \simeq \sqrt{\left(\frac{1}{Q} \frac{\partial
Q}{\partial a}\right)^2 s^2_{<a>} + \left(\frac{1}{Q}
\frac{\partial Q}{\partial b}\right)^2 s^2_{<b>}}.
\end{equation}

The power of the Kolmogorov-Smirnov test has already been proven in the
field of 
astronomy, even multidimensional versions of the Kolmogorov-Smirnov test are
developed for astronomical purposes (Fasano \& Fran\-ce\-schi\-ni 1987;
Gosset 1987). Fasano \& Franceschini (1987) demonstrate even that the
Kolmogorov-Smirnov test is a much more goodness-of-fit test than the often used
$\chi^2$-test. It is, however, the first time that this test is used to evaluate
synthetic spectra with respect to observational data. 

To summarize the strategy: \\
1) from the large set of stars and two high-resolution observations, it is
possible to indicate calibration problems (see forthcoming paper);\\
2) in the comparison between observed and synthetic spectrum, some discrepancies
were clearly visible. The knowledge on the relative importance of the different
molecules (Fig. \ref{atdiv}) and on the influence of the various stellar
parameters (Fig. \ref{overall}) makes it possible to elicit the origin of these
discrepancies. This is due to the presence of different molecules in this large
wavelength range, whose spectral shape and strength are a very good diagnostic
tool to reveal the stellar parameters;\\
3) once a high level of accuracy is achieved, the Kol\-mo\-go\-rov-Smirnov test can be
used for further refinement. From the ($Y-F$) plots, one can deduce systematic
errors, yielding an indication on inaccurate stellar parameters. By using the
results illustrated in Fig. \ref{overall}, one can improve the stellar
parameters. The stellar parameters, used to generate a synthetic spectrum, are
taken to represent the `true' stellar parameters, once the $\beta$-values are
lower than a certain maximum value.\\

\section{Discussion}

\begin{figure*}[t!]
\begin{center}
\scalebox{0.65}[0.65]{\rotatebox{90}{\includegraphics{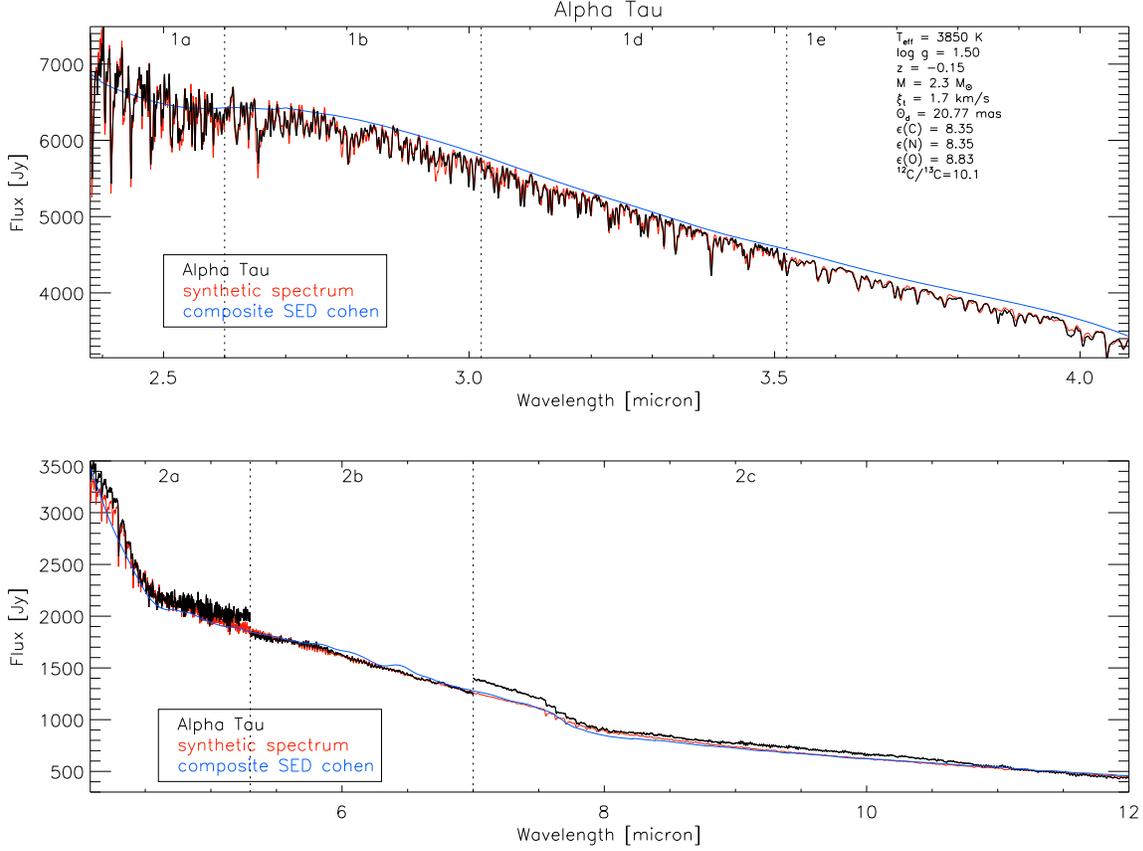}}}
\caption{Comparison between band 1 and band 2 of the ISO-SWS data
of $\alpha$ Tau (black) and the synthetic spectrum (red) with
stellar parameters \teff\ = 3850 K, $\log$ g = 1.50, M = 2.3 \Msun,
z = $-0.15$, \vt\ = 1.7 \km, \cc\ = 10,
$\varepsilon$(C) = 8.35, $\varepsilon$(N) = 8.20, $\varepsilon$(O)
= 8.83 and \ad\ = 20.77 mas. The composite SED of Cohen et al.
(1992) is plotted in blue. \label{result}}
\end{center}
\end{figure*}

Our best fit involves a synthetic spectrum with stellar parameters
\teff\ = 3850 K, $\log$ g = 1.50, M = 2.3 \Msun, z = $-0.15$, \vt\ =
1.7 \km, \cc\ = 10, $\varepsilon$(C) = 8.35,
$\varepsilon$(N) = 8.35, $\varepsilon$(O) = 8.83 and \ad\ = 20.77
mas (see Fig. \ref{result}). The uncertainties are estimated to be
$\Delta$\teff\ $= 70$ K, $\Delta\log$ g$ = 0.15$, $\Delta$\vt\ $= 0.3$ \km,
$\Delta$z $= 0.20$, $\Delta\varepsilon$(C) $= 0.20$,
$\Delta\varepsilon$(N) $= 0.25$, $\Delta\varepsilon$(O) $= 0.15$,
$\Delta$\cc\ $= 1$ and $\Delta\theta_D = 0.03$. The quoted
accuracy of 0.03 mas does not take the absolute photometric flux
accuracy into account, which is 5\,\% in band 1, resulting in
$\Delta\theta_D < 0.83$ mas.

Due to the minor influence of the stellar mass, the uncertainty on this
parameter, which follows from the
construction of a synthetic spectrum, is quite large. But by using the
parallax this uncertainty is constrained to $\Delta$M $= 0.8$ \Msun. The
parallax measurement by Hipparcos for $\alpha$ Tau is $50.09 \pm 0.95$ mas,
which corresponds to a distance of $19.96 \pm 0.39$ pc. With an angular
diameter \ad\ of $20.77 \pm 0.83$ mas, a radius R $ = 44.6
\pm 2.0$ R$_{\odot}$ is found, which combined with $\log$ g $ = 1.50 \pm
0.15$, implies a mass of $2.3 \pm 0.8$ \Msun. A refinement in the mass
determination was proposed by El Eid (1994), who noted a
correlation between the $^{16}$O/$^{17}$O ratio and stellar mass
for evolved stars. Using this ratio in conjunction with
evolutionary calculations, El Eid (1994) found a mass of 1.5
\Msun\ for $\alpha$ Tau. A comparison of the deduced luminosity
and effective temperature with evolutionary tracks (cfr.
McWilliam 1990) indicates a mass of 2.0 \Msun\ for $\alpha$ Tau.

The accuracy of the method can be tested by making a grid around
the resultant final stellar parameters. This is illustrated for
$\alpha$ Tau in Table \ref{grid}. The $\beta$-values, corresponding to the final
stellar paramters, for the different sub-bands are
$\beta_{\mathrm{1A}} = 0.040$, $\beta_{\mathrm{1B}} = 0.034$,
$\beta_{\mathrm{1D}} = 0.036$, $\beta_{\mathrm{1E}} = 0.027$ (see
Table \ref{beta}). In Table \ref{grid}, the $\beta$-values are
given for a change of one of the main stellar parameters, being
the temperature, the gravity, the metallicity, the carbon
abundance, the nitrogen abundance and the oxygen abundance. From this
table, one can 
deduce that there are four other acceptable sets of parameters,
indicated by a grey box. The final stellar parameters --- as
mentioned above --- were chosen to represent the `true' atmospheric
parameters of $\alpha$ Tau, due to the lower $\beta$-value in band
1A, where the important molecule CO is absorbing.

\begin{table}[h!]
\caption{\label{grid}$\beta$-values from the Kolmogorov-Smirnov test for the
evaluation of several synthetic spectra w.r.t. the ISO-SWS spectrum of $\alpha$
Tau. If not specified, the used stellar parameters were \teff\ = 3850
K, $\log$ g = 1.50, \vt\ = 1.7 \km, [Fe/H] = $-0.15$,
$\varepsilon$(C) = 8.35, $\varepsilon$(N) = 8.35, $\varepsilon$(O)
= 8.83, \ad\ = 20.77. The $\beta$-values for these stellar parameters are given
as $\beta_{\mathrm{final}}$. } \vspace{1ex}
\begin{center}
\setlength{\tabcolsep}{0mm}
\begin{tabular}{r||cccc}  \hline
\rule[0mm]{0mm}{5mm} & $\beta_{\mathrm{1A}}$ &
$\beta_{\mathrm{1B}}$ & $\beta_{\mathrm{1D}}$ &
$\beta_{\mathrm{1E}}$ \\ $\beta_{\mathrm{max}}\ \ $ & 0.060 &
0.050 & 0.040 & 0.040 \\
\rule[-3mm]{0mm}{3mm}$\beta_{\mathrm{final}}$\ \  \ & 0.040 &
0.034 & 0.036 & 0.027 \\ \hline \rule[0mm]{0mm}{5mm}\teff\ = 3710\
\ \  & \ \ 0.080\ \  & \ \ 0.050\ \  & \ \ 0.059\ \  & \ \ 0.027\
\ \\ \colorbox[gray]{0.85}{3780\ } & \colorbox[gray]{0.85}{\ \
0.054\ \ } & \colorbox[gray]{0.85}{\ \ 0.028\ \ } &
\colorbox[gray]{0.85}{\ \  0.026\ \ } & \colorbox[gray]{0.85}{\ \
0.017\ }\ \\ 3920 \ \  & 0.031 & 0.043 & 0.084 & 0.028 \\
\rule[-3mm]{0mm}{3mm}3990 \ \  & \ \ 0.044 & 0.046 & 0.104 & 0.041
\\ \hline \rule[0mm]{0mm}{5mm}$\log$ g = 1.20 \ \  & 0.037 & 0.093
& 0.071 & 0.015 \\ 1.35 \ \  & 0.034 & 0.063 & 0.064 & 0.018 \\
\colorbox[gray]{0.85}{1.65\,} & \colorbox[gray]{0.85}{\ \ 0.045\ \
} & \colorbox[gray]{0.85}{\ \ 0.028\ \ } & \colorbox[gray]{0.85}{\
\ 0.044\ \ } & \colorbox[gray]{0.85}{\ \ 0.027\ }\ \\
\rule[-3mm]{0mm}{3mm}1.80 \ \  & 0.051 & 0.055 & 0.032 & 0.027\\
\hline \rule[0mm]{0mm}{5mm}z = $-0.45$ \ \  & 0.037 & 0.094 &
0.075 & 0.013\\ $-0.30$ \ \  & 0.034 & 0.062 & 0.066 & 0.015\\
\colorbox[gray]{0.85}{0.00\,} & \colorbox[gray]{0.85}{\ \ 0.045\ \
} & \colorbox[gray]{0.85}{\ \ 0.028\ \ } & \colorbox[gray]{0.85}{\
\ 0.043\ \ } & \colorbox[gray]{0.85}{\ \ 0.024\ }\ \\
\rule[-3mm]{0mm}{3mm}0.15 \ \  & 0.050 & 0.054 & 0.029 & 0.030 \\
\hline \rule[0mm]{0mm}{5mm}$\varepsilon$(C) = 8.05 \ \  & 0.047 &
0.132 & 0.040 & 0.018\\ 8.20 \ \  & 0.042 & 0.065 & 0.045 &
0.021\\ 8.50 \ \  & 0.036 & 0.093 & 0.063 & 0.031 \\
\rule[-3mm]{0mm}{3mm}8.65 \ \  & 0.065 & 0.174 & 0.085 & 0.044 \\
\hline \rule[0mm]{0mm}{5mm}$\varepsilon$(N) = 7.85 \ \ & 0.043 &
0.032 & 0.051 & 0.019 \\ 8.10 \ \ & 0.041 & 0.031 & 0.051 & 0.021 \\ 8.60 \ \ & 0.034
& 0.032 & 0.055 & 0.025 \\ \rule[-3mm]{0mm}{3mm}8.85 \ \ & 0.031 & 0.034 & 0.059 &
0.028 \\  
\hline \rule[0mm]{0mm}{5mm}$\varepsilon$(O) = 8.63 \ \  & 0.057 &
0.049 & 0.096 & 0.046 \\ 8.73 \ \  & 0.035 & 0.040 & 0.075 & 0.034
\\ \colorbox[gray]{0.85}{8.93\,}& \colorbox[gray]{0.85}{\ \ 0.052\
\  } & \colorbox[gray]{0.85}{\ \ 0.019\ \  } &
\colorbox[gray]{0.85}{\ \ 0.026\ \  } & \colorbox[gray]{0.85}{\ \
0.018\ }\ \\ \rule[-3mm]{0mm}{3mm}8.03 \ \  & 0.070 & 0.043 &
0.028 & 0.019 \\ \hline
\end{tabular}
\end{center}
\end{table}

According to Fig. 20 in Bessell et al.\ (1998), $\alpha$
Tau, with a ($V-K$) colour index of 3.67, has a bolometric
correction BC$_K$ of $2.60 \pm 0.05$. From the bolometric magnitude
$m_{\mathrm{bol}} = -0.21 \pm 0.05$, the distance $d = 19.964 \pm 0.379$, the
assumption that the interstellar extinction may be neglected, and by adapting
the absolute bolometric magnitude of
the Sun $M_{\mathrm{bol},\odot} = 4.74$ (according to Bessell et al. 1998),
one deduces the luminosity of $\alpha$
Tau L $ = 382 \pm 23$ L$_{\odot}$. A radius R of $44.6 \pm 2.4$ \Rsun\
(calculated from the distance $d$ and the angular diameter \ad)
yields an effective temperature of $3828 \pm 180$ K.
This is in very good agreement with the effective temperature deduced from the
ISO-SWS spectrum, being \teff\ $= 3850 \pm 70$ K. This latter \teff, in
conjunction with L $ = 382 $L$_{\odot}$, would give R $ = 44.05 \pm 5.55$ \Rsun\
and \ad\ $= 20.53 \pm 2.6$ mas.

It is clear that a gravity of $\log$ g = $1.50 \pm 0.15$ dex agrees better with a
`physical' gravity ($\log$ g$ = 1.44$ from M = 2 \Msun\ and R = 44.6 \Rsun)
determined by the mass M, the radius R --- or the effective temperature \teff\
and the luminosity L --- and/or evolutionary tracks than with a `spectroscopic'
gravity of e.g.\ Lambert \& Ries (1981), Kov\'{a}cs (1983) and Luck \&
Challener (1995) (see Appendix A). It is well known that the accuracy of
spectroscopic determinations of gravities from ionization equilibria or
molecular equilibria for individual stars is not very good for red giants (cf.,
e.g., Trimble \& Bell 1981, Brown et al. 1983, Smith \& Lambert
1985). Smith \& Lambert (1985) suggested that the low spectroscopic gravity
$\log$ g = 0.8 may be due to non-LTE effects on iron lines. After corrections
for 
these effects, Smith \& Lambert (1990) finally determined values for the
various atmospheric parameters, with which our results agree very well. Our
method for gravity determination from infrared spectra appears to be much less
affected by NLTE effects, being based on a broad range of diagnostics over a
large spectral interval. A great advantage of the used Kolmogorov-Smirnov
test, is also that several deviation estimating parameters $\beta$ are
calculated, which are not independent and are each of them related to the
specific pattern of behaviour of absorption of the different molecules. E.g.\
the abundance of carbon is mainly determined from the band-1A and band-1B data,
while a 
wrong gravity is traced from the $\beta_{\mathrm{1B,1D}}$-values. Simulations
with other ($\log$ g, $\varepsilon$(C))-combinations never reached the same
level of agreement as for the ($\log$ g = 1.50, $\varepsilon$(C) =
8.35)-combination. Using e.g.\ $\log$ g = 1.35 requires a lower carbon abundance
(than $\varepsilon$(C) = 8.35) in order to fit the CO line strengths. The best
fit then is obtained for $\varepsilon$(C) = 8.25, with $\beta_{\mathrm{1A}}$ =
0.033 and $\beta_{\mathrm{1B}}$ = 0.024. The corresponding
$\beta_{\mathrm{1D}}$-value is then, however, 0.056 arising from too steep a
continuum.

The mass M, luminosity L and radius R are the three fundamental
parameters of a (spherical) star where two of these may be
replaced by the surface gravity g = $G $M/R$^2$ and the effective
temperature $\sigma$\teff$^4$=L/(4$\pi$R$^2$). Both the mass and
the luminosity are measurable physical quantities, whereas the
radius of a star is a fictitious quantity because a star is a
gaseous sphere and does not have a sharp edge. The relevant
observable quantity is the center-to-limb variation (=clv) of
intensity or limb darkening. The sun is the only star whose clv
one can observe directly. Direct diameter determinations are based
on a standard procedure which evaluates interferometric, lunar
occultation or binary eclipse data by determining a radius
position on the base of a parameterized approximation (e.g.\
uniform disk) of a model-predicted limb-darkening curve (see e.g.\
Scholz 1998). The radius is then taken at $\mu = 0$ ($\mu =
\cos\theta$, with $\theta$ the angle between the radius vector and
the line of sight). Published limb-darkening coefficients (for an
overview, see Scholz (1998)) are then needed to convert to the
limb-darkened corrected angular diameter. In Table
\ref{literature}, this limb-darkened angular diameter is listed
for the direct angular diameter determinations, except when
otherwise mentioned. The angular diameter value derived by us from the
ISO-SWS data is however a `theoretical' radius deduced from an indirect
method. It corresponds to the definition
$\sigma$\teff$^4$ = L/(4$\pi$R($\tau_{\mathrm{ross}}=1$)$^2$) and $\log$
g = GM/R($\tau_{\mathrm{ross}}=1$)$^2$ (Plez et al.\ 1992).
Baschek et al.\ (1991) have summarized various radius
definitions. One of the most common theoretical definitions for
the radius is the {\it{optical depth radius}}. Assuming that most
of the flux emerges from a layer with optical depth $\tau \approx
1$, is the base for defining a radius R$_{\lambda}$ =
r($\tau_{\lambda} = 1$), which depends of course on the extinction
coefficient at this wavelength. In Fig. 2 of Scholz (1998) one can
see clearly that there is no trivial correlation between
the clv shape and the position of this $\tau_{\lambda} = 1$
radius. The angular diameter (and radius) obtained from the
ISO-SWS data should therefore be only compared to other indirect
spectrophotometric angular diameter values, resulting from e.g.\
comparing the observed flux at a distance d from the star with the
surface flux, the surface brightness method (introduced by Barnes
\& Evans 1976) or the infrared or Rayleigh-Jeans flux method (from
Blackwell \& Shallis 1977). The quadratic dependence of the
observed flux on the angular diameter makes an accurate
determination of \ad\ from the ISO-SWS data possible. With the exception of a
few
references given in the catalogue of Fracassini et al.\ (1988), the
angular diameter \ad\ $ = 20.77 \pm 1.07$ mas is in very good
agreement with the other indirect determined angular diameter
values.

The carbon abundance is set by the strength of the CO absorption,
while the strength of the OH lines yields the oxygen abundance.
Due to the weakness of the NH and CN lines in $\alpha$ Tau (see
Fig. \ref{atdiv}), it was impossible to determine $\varepsilon$(N) directly from
the SWS spectrum. Since the chemical equilibrium detemines however the shape of
the synthetic 
spectrum, a too inaccurate nitrogen abundance would result in high
$\beta$-values, which did not happen when using $\varepsilon$(N) = 8.35 (see
Table \ref{grid}).
The large dependence of the CO line-width on $\varepsilon$(C) implies an
uncertainty $\Delta \varepsilon(\mathrm{C}) < 0.10$~dex. This can also be seen
in Table \ref{grid}: changing the carbon abundance by 0.15 dex results in e.g.\
too high a $\beta_{\mathrm{1B}}$-value so that only for
$\varepsilon(\mathrm{C})$ no other acceptable set of $\beta$-values appears in
Table \ref{grid}. Due to, however, the larger standard deviation after rebinning
in band 1A and the beginning of band 1B than in the other sub-bands, the total
uncertainty on 
$\varepsilon(\mathrm{C})$ is taken to be 0.20~dex. From Table \ref{grid}, one
also can deduce that the error bars on $\varepsilon$(N) and $\varepsilon$(O),
being 0.25~dex and 0.15~dex respectively, are realistic values.
The carbon, nitrogen and oxygen abundance, together with the low isotopic
\cc\ ratio of $\alpha$ Tau are an argument for $\alpha$ Tau
being on the first giant branch.

With a luminosity of 382 L$_{\odot}$ and \teff\ $ = 3850$ K, $\alpha$ Tau lies
slightly above the position where core helium-burning starts for stars with 1.0
\Msun\ $\le$ M $\le$ 2.0 \Msun\ and has an age of $\approx$ 5 Gyr in the
evolutionary diagram of Claret \& Gimenez (1995). Consequently
there are two
possibilities for the evolutionary stage of $\alpha$ Tau: \\
(i) $\alpha$ Tau is still on the first giant branch, or\\
\ (ii) $\alpha$ Tau has ignited helium-burning and either had not enough time to
move into the region of the clump stars or is already on the asymptotic giant
branch. \\
Following the discussion of Kov\'{a}cs (1983), it seems impossible to
decide whether $\alpha$ Tau is in phase (i) or in phase (ii).
So far, the low \cc\ ($= 10$) value cannot be explained by standard
evolutionary models for low-mass stars. Charbonnel (1994) introduced
an extra-mixing process that occurs later on the giant branch and produces an
additional decrease of the surface \cc\ and C/N values. This extra-mixing
process should be only efficient when the hydrogen-burning shell has reached the
discontinuity in molecular weight. Charbonnel (1994) held the process of Zahn
(1992) --- who proposed a consistent picture of the interaction between
meridional circulation and turbulence induced by rotation in stars ---
responsible for the required extra-mixing on the red giant branch of low mass
stars. 

We want to stress that the current stellar parameters are obtained by using
the 1-dimensional models as described in Sect.\ \ref{models}. Since there are
indications of problems 
with the temperature distribution in the outermost layers of the theoretical
models (see forthcoming paper), a future project is to extend the theoretical
models to 
include the ability to conduct temperature perturbations in order to simulate
convection, the presence of a chromosphere or a dust shell and a change in
opacity. Using these improved models, it will be interesting to study the
changes in the abundance (especially carbon, nitrogen and oxygen) needed to
match the line strengths of CO (fundamental and first overtone)lines, SiO
(fundamental and first overtone) lines, H$_2$O fundamental lines and OH
fundamental lines in the ISO-SWS spectra and high-resolution Fourier Transform
Spectrometer (FTS) spectra. In addition, by studying the effect of changing the
microturbulence \vt, it might be possible to identify a typical region
associated with a specific spectral type in the 3-dimensional ($\Delta T$, \vt,
$\varepsilon$)-plane, with $\varepsilon$ being the abundance of carbon, nitrogen
of oxygen. So far, it was still not possible to identify such a region from the
many different values for the stellar parameters of red giants derived from
different methods and/or data (cf., e.g., Lambert \& Ries (1981) versus
McWilliam (1990)). Only a systematic study may reveal such a region.

The composite SED of Cohen et al.\ (1992) is also plotted on Fig. \ref{result}
and lies $\sim 2$\,\% higher than the SWS data, which is well within the quoted
flux calibration accuracy. Cohen et al.\ (1992) noted however a global accuracy of $\sim
1.5$\,\% for this $\alpha$ Tau SED and a local accuracy which is somewhat
lower. The accuracy quoted by Cohen et al.\ (1992) thus appears slightly
optimistic.

The remaining discrepancies seen in band 2 are due to the memory effects
which affect the reliability of observations in this band. The obtained
synthetic spectrum will be used as a test vehicle for the memory correction
procedures which are currently under development.

\section{Conclusions}

The full scan AOT01 speed-4 ISO-SWS observation of $\alpha$ Tau
has been used to determine the stellar parameters of this K5
giant. Due to the complementarity of the way parameters affect the
spectrum in the wavelength range from 2.38 -- 12 \mic\ --- like the
presence of different molecules which are each dependent in
another way on the several stellar parameters --- it was possible to
pin down the atmospheric parameters with a high accuracy. This
resulted in the following parameters \teff\ = $3850 \pm 70$ K,
$\log$ g = $1.50 \pm 0.15$ , M = $2.3 \pm 0.8$ \Msun, z = $-0.15
\pm 0.20$, \vt\ = $1.7 \pm 0.3$ \km, \cc\ = $10 \pm 1$,
$\varepsilon$(C) = $8.35 \pm 0.20$, $\varepsilon$(N) = $8.35 \pm
0.25$, $\varepsilon$(O) = $8.83 \pm 0.15$ and \ad\ = $20.77 \pm
0.83$ mas. These parameters were compared with and discussed
w.r.t. the results listed by other authors, one of the most
striking result being that one should rely more upon a physically
determined gravity than upon a spectroscopic gravity. The good consistency
between the stellar parameters deduced from the ISO-SWS data, colours,
high-resolution (optical) spectra, IRFM, ... proves that it is possible to
extract good quantative informations from these intermediate-resolution ISO-SWS
spectra! 
 
\appendix
\section{Comments on the different stellar parameters of $\alpha$ Tau.}

1. Mc William (1990) based his results on high-resolution
spectroscopic observations with resolving power 40000. \\The
effective temperature was determined by empirical and
semi-empirical results found in the literature and from broad-band
Johnson colours. The gravity was ascertained by using the
well-known relation between g, \teff, the mass M and the
luminosity L, where the mass was determined by locating the stars
on theoretical evolutionary tracks. So, the computed gravity is
fairly insensitive to errors in the adopted L. High-excitation
iron lines were used for the metallicity [Fe/H] in order that the results are
less spoiled by non-LTE effects. The author refrained from determining the
gravity
in a spectroscopic way (i.e.\ by requiring that the abundance of
neutral and ionized species yield the same abundance) because
{\it{`a gravity adopted by demanding that neutral and ionized
lines give the same abundance, is known to yield temperatures
which are $\sim 200$ K higher than found by other methods. This
difference is thought to be due to non-LTE effects in Fe~I
lines.'}} By requiring that the derived iron abundances, relative
to the standard 72 Cyg, were independent of equivalent width, the
microturbulent velocity $\xi_t$ was found.

2. Lambert \& Ries (1981) have used high-resolution low-noise
spectra. The parameters were ascertained by demanding that the
spectroscopic requirements (ionization balance, independence of
the abundance of an ion versus the excitation potential and
equivalent width) should be fulfilled.
 The effective temperature
was found from the Fe~I excitation temperature and the model
atmosphere calibration of the excitation temperature as a function
of \teff. As quoted by Ries (1981), Harris et al.\ (1988) and Luck
\& Challener (1995) their \teff\ and $\log$ g are too high and
should be lowered by 240 K and 0.40 dex, respectively. The isotopic
ratio \cc\ was taken from Tomkin et al.\ (1975), while the
luminosity was estimated from the K-line visual magnitude $M_V(K)$ given by
Wilson (1976) and the bolometric correction BC by Gustafsson \& Bell (1979). The
abundances of carbon, nitrogen and oxygen were 
based on C$_2$, [O~I] and the red system CN lines respectively.
Luck \& Challener (1995) wondered whether the nitrogen
abundance [N/Fe] = $-0.20$ quoted reflects a typographic error and should rather
be 
[N/Fe]=+0.20, resulting in $\varepsilon$(N)=7.86, which is more in
agreement with being a red giant branch star.

3. The effective temperature \teff\ and the angular diameter \ad\ given by
Blackwell et al.\ (1990) were determined by the infrared flux
method (IRFM), a semi-empirical method which relies upon a
theoretical calibration of infrared bolometric corrections with
effective temperature. One expects that the IRFM should yield results better
than 1\,\% for the effective temperature and 2 -- 3\,\% for the angular
diameter. The final effective temperature is a weighted mean of
T(J$_n$), T(K$_n$) and T(L$_n$), with J$_n$ at 1.2467 \mic, K$_n$
at 2.2135 \mic\ and L$_n$ at 3.7825 \mic.

4. -- 5. The stellar parameters quoted by Tsuji (1986, 1991) were based on
the results of Tsuji (1981), in which the temperature was
determined by the IRFM method. A mass of 3 \Msun\ was assumed to
ascertain the gravity. Tsuji (1986) has used high-resolution FTS
spectra of the CO (first overtone) lines to determine
$\varepsilon$(C) and \vt\ by assuming that the abundance should be
independent of the equivalent width of the lines. In Tsuji (1991)
CO lines of the second overtone were used, where a standard
analysis yielded results of $\varepsilon$(C) = 8.39 and a linear
analysis of weak lines led to $\varepsilon$(C) = 8.31.

6. High-resolution spectra of OH ($\Delta\nu=2$), CO($\Delta\nu=3$),
CO($\Delta\nu=2$) and CN($\Delta\nu=2: A ^2\Pi-X ^2\Sigma$) were obtained by
Smith \& Lambert (1985). They have used ($V-K$) colours and the calibration
provided by Ridgway et al.\ (1980) to determine \teff. Using the spectroscopic
requirement that $\varepsilon$(Fe~I) = $\varepsilon$(Fe~II) yields $\log$ g =
0.8 dex, which is too low for a K5III giant. They suggested that the reason for
this low value is the overionization of iron relative to the LTE situation. They
then have computed the surface gravity by using a mass estimated from
evolutionary tracks in the H-R diagram. The metallicity was taken from
Kov\'{a}cs (1983) and for the microturbulence they used Fe~I, Ni~I and Ti~I
lines, demanding that the abundances are independent of the equivalent
width. Using the molecular lines, they determined $\varepsilon$(C),
$\varepsilon$(N), $\varepsilon$(O) and \cc.

7. Harris \& Lambert (1984) have taken \teff, $\log$ g and \vt\
determined by Dominy, Hinkle \& Lambert in 1984, a reference which we could not
trace back. The isotopic ratio \cc\ was adopted from Tomkin et al.\ (1975). The
carbon 
abundance was found by fitting weak $^{12}$C$^{16}$O lines at 1.6, 2.3 and 5
\mic.

8. Kov\'{a}cs (1983) obtained observations at the 1.52\,m telescope of ESO at La
Silla. By using different IR colour indices ($V-R$, $V-I$, $V-J$, $V-K$, $V-L$,
$R-I$ for the broad-band photometry and $52-56$, $52-63$, $52-72$, $52-80$,
$52-86$ for the narrow-band photometry) the effective temperature was found. The
gravity and microturbulent velocity were determined in a spectroscopic way,
where the abundance derived from strong and medium lines should equal the
abundance derived from weak lines for the right value of \vt. The \cc\ ratio was
taken from Tomkin et
al. (1976). Using a parallax of $0.''050$ and \ad\ of 24 mas, resulted in a
radius of $49 \pm 4$ R$_{\odot}$, which corresponds to a luminosity L of $463
\pm 59$ L$_{\odot}$, while $M_V = -0.7$ and a bolometric correction taken from
Johnson (1966) yields L = 413 L$_{\odot}$.

9. Lambert et al.\ (1980) took model parameters from published
papers which constrained the \cc\ ratio. The parameters for $\alpha$ Tau
were based on Tomkin et al.\ (1976) and Lambert (1976). Tomkin et
al. (1976) have used IR colours for the determination of \teff\ and the
microturbulence was ascertained by fitting the theoretical curve of
growth to the $^{12}$CN curve of growth ($^{12}$CN(2-0) around
8000 \ang , weak $^{12}$CN(4-0) lines around 6300\ang\ and weak
$^{12}$CN(4-2) lines around 8430 \ang). Together with $^{13}$CN
lines around 8000\ang, the isotopic ratio \cc\ computed.
The gravities were estimated from the effective temperature, the mass and the
luminosity.

10. By fitting the Engelke function (Engelke 1992) to the template of $\alpha$
Tau, Cohen et al.\ (1996b) could determine  \teff\ and \ad. A gravity of 2.0
was adopted, though in Cohen et al.\ (1992) a value of 1.5 --- taken from Smith
\& Lambert (1985)---  was used.

11. By taking the mean value of the different temperatures derived by
calibrations with photometric indices ($U-B$, $B-V$, $V-R$, $V-I$, $V-J$, $V-K$,
$V-L$)
Fern\'{a}ndez-Villaca\~{n}as et al.\ (1990) have fixed the effective temperature.
For the gravity, the DDO photometry indices $C(45-48)$ and $C(42-45)$ were
used. The Fe~I lines served for the determination of \vt.

12. van Paradijs \& Meurs (1974) have adopted the angular diameter from Currie
et al. (1974). To determine the effective temperature they used several
continuum data, the curve of growth with the line strengths of Fe~I lines and
the surface brightness. Requiring that the neutral and ionized lines of Fe, Cr,
V, Ti and Sc gave the same abundance, yielded the gravity. Gravity, parallax and
angular diameter yielded the mass, while the luminosity was determined from the
effective temperature, the parallax and the angular diameter. van Paradijs \&
Meurs (1974) quoted that Wilson (1972) found [Fe/H] = $-0.69$.

13. Tomkin \& Lambert (1974) obtained photoelectric scans of the red CN lines
with a resolution of 0.05 \ang\ for the red scans and 0.09 \ang\ for the
infrared scans. The effective temperature and gravity were adopted from Conti et
al.
(1967), who have ascertained the effective temperature from the absolute
magnitude derived from the K emission-line width and the gravity from the
effective temperature, the mass and the luminosity. The microturbulent velocity
was determined using a curve of growth technique and the \cc\ ratio was
obtained directly from the horizontal shifts between the curves of growth for
$^{12}$CN and $^{13}$CN lines.

14. Also Luck \& Challener (1995) have determined the effective
temperature using photometric data (DDO $C(42-45)$ and $C(45-48)$,
Geneva $B_2-V_1$, Johnson $V-K$, $J-K$, $V-R$ and $B-V$). Two methods were used
to fix the surface gravity. The first one determined
the `physical gravity' by using the mass M and the radius R, with the mass M
determined by \teff, L($M_V(K)$, BC) and theoretical evolutionary
tracks and the radius R by \teff\ and L($M_V(K)$, BC). The
`spectroscopic gravity' is based on the ionization balance of Fe~I
and Fe~II lines. The difference between these two gravities was
very large, being 1.0 dex! When the Fe~II oscillator strengths
were modified to reflect a solar Fe abundance of 7.50 (instead of
the used 7.67), then the spectroscopic gravity scale would rise by
+0.25 dex (still far less than 1.0 dex) resulting in an increase
of [Fe/H] of $+0.15$ dex. They have quoted both pluses and minuses
for the spectroscopic gravity and a single plus for the physical
gravity. The microturbulence was ascertained by forcing no
dependence of abundance (derived from individual Fe~I lines) upon
the equivalent width. In determining the carbon abundance using the
C$_2$ Swan system lines and [C~I] lines, the nitrogen abundance
and isotopic ratio \cc\ using CN lines and the oxygen abundance
from [O~I] lines, they always found a better trend with
temperature when the spectroscopic gravity was used and also a
better agreement between the two carbon indicators. When comparing
their results with the ones of Lambert \& Ries (1981), they found that $\alpha$
Tau was always the most discrepant star. Their very low
spectroscopic gravity, compared with the results of other authors
using different methods and taking into account possible
non-LTE effects, may be an indication that a value of $\log$ g
$\sim$ 1.5 is in better agreement with the real gravity of
$\alpha$ Tau.

15. Aoki \& Tsuji (1997) took the same stellar parameters as Tsuji (1986,
1991), but they now used CN-lines to determine $\varepsilon$(N) and
$\xi_t$.

16. Bonnell \& Bell (1993) constructed a grid based on  the
effective temperature \teff\ of Manduca et al.\ (1981). They
used ground-based high-resolution FTS spectra of OH and [O~I]
lines. The requirement that the oxygen abundances determined from
the [O~I] and OH line widths agree amounts to finding the
intersection of the loci of points defined by the measured widths
in the ([O/H], $\log$ g) plane. For $\alpha$ Tau they found large
discrepancies in the [O~I] oxygen abundance, which leads to a
spread of $+0.3$ dex in $\log$ g. The determination of \vt\ was based on the
OH-lines, but was hampered by a lack of weak lines from this radical.
Using a least square fit, they found
$\xi_t = 1.05$ \km\ for the OH($\Delta\nu = 2$) lines and $\xi_t =
1.62$ \km\ for the OH($\Delta\nu = 1$) lines, reflecting that the
OH($\Delta\nu = 2$) sequence lines are formed at greater average
depth relative to the OH($\Delta\nu = 1$) lines. For the
[O~I] lines a microturbulent velocity of 1.5 \km\ or 2.0 \km\ was
used.

17. Ridgway et al.\ (1982) determined the limb-darke\-ning-corrected angular
diameter $\theta_{\mathrm{LD}} = \theta_d$ using the lunar occultation technique.

18a--b. Di Benedetto \& Rabbia(1987) used Michelson interferometry by the
two-telescope baseline located at CERGA. Combining this angular diameter with
the bolometric flux
$F_{\mathrm{bol}}$ (resulting from a directed integration using the trapezoidal
rule over the flux distribution curves, after taking  interstellar absorption
into account) they \\found an effective temperature of 3970 K, which is in good
agreement with the results obtained from the lunar occultation technique.
Di Benedetto (1998) calibrated the surface brightness-colour correlation
using a set of high-precision angular diameters measured by modern
interferometric techniques. The stellar sizes predicted by this correlation were
then combined with bolometric flux measurements, in order to determine
one-dimensional (T, V-K) temperature scales of dwarfs and giants.

19. Mozurkewich et al.\ (1991) used the MarkIII Optical
Interferometer. The uniform-disk angular diameter $\theta_{\mathrm{UD}}$ had a
residual of 1\,\% for the 800 nm observations and less than 3\,\% for the 450 nm
observations. The limb-darkened
diameter was then obtained by multiplying the uniform-disk angular diameter with
a correction factor (using the quadratic limb-darkening coefficient from
Manduca 1979).

20. Quirrenbach et al.\ (1993) have determined the u\-ni\-form-disk angular diameter
in the strong TiO band at 712 nm and in a continuum band at 754 nm with the
MarkIII stellar interferometer on Mount Wilson. Because limb darkening is
expected to be substantially larger in the visible than in the infrared, the
measured uniform-disk diameters should be larger in the visible than in the
infrared. This seems, however, not always to be the case. Using the same factor
as Mozurkewich et al.\ (1991) we have converted their continuum uniform-disk
value (19.80 mas) into a
limb-darkened angular diameter, yielding a value of 22.73 mas, with a
systematic uncertainty in the limb-darkened angular diameter of the order of
1\,\% in addition to the measurement uncertainty of the uniform-disk angular
diameter (Davis 1997).

21. Volk \& Cohen (1989) mentioned a distance of $20.0 \pm 2.0$ pc. The
effective temperature was directly determined from the literature values of
angular diameter measurements and total flux observations (also from
literature). The distance was taken from the Catalogue of Nearby Stars (Gliese,
1969) or from the Bright Star Catalogue (Hoffleit 1982).

22. Blackwell et al.\ (1991) is a revision of Blackwell et al.\ (1990) where the
H$^-$ opacity has been improved. They investigated the effect of the
improved
H$^-$ opacity on the IRFM temperature scale and derived angular diameters. Also
here, the mean temperature is a weighted mean of the temperatures for $J_n$,
$K_n$ and $L_n$. Relative to Blackwell et al.\ (1990) there was a change of
temperature up to 1.4\,\% and an decrease by 3.5\,\% in \ad.

23. By using a \teff-($V-I$) transformation of Johnson (1966), Linsky \& Ayres
(1978) have determined the effective temperature.

24. Bell \& Gustafsson (1989) first determined the temperature from the
Johnson K band at 2.2 \mic\ by use of the IRFM method. By comparison with
temperatures deduced from the colours Glass $J-H$, $H-K$, $K-L$ and $K$; Cohen,
Frogel and Persson $J-H$, $H-K$, $K-L$ and $K$; Johnson $V-J$, $V-K$, $V-L$ and
$K$; Cousins $V-R$, $R-I$; Johnson and Mitchell 13-colour and Wing's
near-infrared eight-colour photometry, they found that \teff(IRFM) was $\sim
80$~K too high, by which they corrected the temperature. The gravity and the
metallicity were adopted from Kov\'{a}cs (1983).

25. Taylor (1999) prepared a catalogue of temperatures and [Fe/H]
averages for evolved G and K giants. This catalogue is available at CDS via
anonymous ftp to cdsarc.u-strasbg.fr

26. Burnashev (1983) has determined \teff, $\log$ g and [Fe/H]
from narrow-band photometric colours in the visible part of the
electromagnetic spectrum.

27. Fracassini et al.\ (1988) have made a catalogue of stellar apparent diameters
and/or absolute radii, listing 12055 diameters for 7255 stars. Only the most
extreme values are listed. References and remarks to the different values of the
angular diameter and radius may be found in this catalogue. Also here these
angular diameter values are given in italic mode when determined from direct
methods and in normal mode for indirect (spectrophotometric) determinations.

28. Perrin et al.\ (1998) have derived the effective temperatures for nine giant
stars from diameter determinations at 2.2 \mic\ with the FLUOR beam combiner on
the IOTA interferometer. This yielded the uniform-disk angular diameter of
$\alpha$ Boo and $\alpha$ Tau of our sample.
The averaging effect of a uniform model leads to an underestimation of the
diameter of 
the star. Therefore, they have fitted their data with limb-darkened disk models
published in the literature. The average result is a ratio between the uniform
and the limb-darkened disk diameters of 1.035 with a dispersion of 0.01.
This ratio could then also be used for the uniform-disk angular diameters of
$\gamma$ Dra and $\beta$ Peg, listed by Di Benedetto \& Rabbia (1987), and $\alpha$
Cet, which was based on a photometric estimate. Several photometric sources were
used to determine the bolometric flux, which then, in conjunction with the
limb-darkened diameter, yielded the effective temperature.

29. Engelke (1992) has derived a two-parameter analytical expression
approximating the long-wavelength (2 -- 60 \mic) infrared continuum of stellar
calibration standards. This generalized result is written in the form of a
Planck function with a brightness temperature that is a function of both
observing wavelength and effective temperature. This function is then fitted to
the best empirical flux data available, providing thus the effective temperature
and the angular diameter.

30. Blackwell \& Shallis (1977) have described the Infrared Flux Method (IRFM)
to determine the stellar angular diameters and effective temperatures from
absolute infrared photometry. For 28 stars (including $\alpha$ Car, $\alpha$
Boo, $\alpha$ Cma, $\alpha$ Lyr, $\beta$ Peg, $\alpha$ Cen A, $\alpha$ Tau and
$\gamma$ Dra) the angular diameters are deduced. Only for the first four stars
the corresponding effective temperatures are computed.

31. Scargle \& Strecker (1979) have compared the observed infrared flux curves
of cool stars with theoretical predictions in order to assess the model
atmospheres and to derive useful stellar parameters. This comparison yielded the
effective temperature (determined from flux curve shape alone) and the angular
diameter (determined from the magnitudes of the fluxes). The overall uncertainty
in \teff\ is probably about 150 K, which translates into about a 9\,\% error in
the angular diameter.

32. Manduca et al.\ (1981) have compared absolute flux measurements in the
2.5 -- 5.5 \mic\ region with fluxes computed for model stellar atmospheres. The
stellar angular diameters obtained from fitting the fluxes at 3.5 \mic\ are in
good agreement with observational values and with angular diameters deduced from
the relation between visual surface brightness and ($V-R$) colour. The
temperatures obtained from the shape of the flux curves are in satisfactory
agreement with other temperature estimates. Since the average error is expected
to be well within 10\,\%, the error for the angular diameter is estimated to be
in the order of 5\,\%.

33. Smith \& Lambert (1990) have determined the chemical composition of a
    sample of M, MS and S giants. The use of a slightly different set of lines
    for the molecular vibra-rotational lines, along with improved $gf$-values
    for CN and NH,  
    lead to a small difference in the carbon, nitrogen and oxygen abundance
    compared to Smith \& Lambert (1985).

\section{The approximate absorption coefficient ${\cal{R}}$}

Kj{\ae}rgaard et al.\ (1982) examined two cases for which they
derived equations for the approximate absorption coefficient
${\cal{R}} = l_{\nu}/ \kappa_{\nu}$. In case (a) most of the carbon and nitrogen
are still considered to be free atoms, assumed for \teff\ $\ge$
5000 K, while in case (b) almost all the C and N are in the form
of CO and N$_2$, assumed for \teff\ $\le$ 4500 K. Looking at the
relatively high partial pressure of CN, it seems that the assumption
of complete association of N into N$_2$ is not valid for our giant
models. Attributing all depletion to the formation of N$_2$ seems
to be more valid in a dwarf model (Bell \& Tripicco 1991). The
formula can however still give a qualitative view of the influence
of the different parameters. For case (b) Kj{\ae}rgaard et al.
deduced an approach for CN:
\begin{eqnarray}
\lefteqn{\mathrm{{\cal{R}}(CN) = }} \nonumber\\
& & \mathrm{(N/H)^{0.5}(C/H)(O/H-C/H)^{-1}(Fe/H)^{-0.1}g^{-0.75}}.
\nonumber \end{eqnarray}
Using the approach of Kj{\ae}rgaard et al., similar equations for
CO, SiO, OH, H$_2$O and NH are derived. 

For CO, we have
\begin{eqnarray}
\lefteqn{\mathrm{{\cal{R}}(CO) \equiv
\frac{\kappa(CO)}{\kappa(H^-_{\mathrm{ff}})}}} \\
 & &  \propto \mathrm{\frac{n(CO)}{n(H^-)}} \\
 & &  = \mathrm{\frac{n(C~I) n(O~I) K(H^-)}{K(CO) n(H~I) n(e)}}.
\end{eqnarray}
In the assumption that all carbon is locked into CO, \\${\cal{R}}$(CO) may also
be written as
\begin{eqnarray}
\lefteqn{\mathrm{{\cal{R}}(CO)}   \propto  \mathrm{\frac{n(C)}{n(H^-)}}}  \\
 & & = \mathrm{\frac{n(C) K(H^-)}{n(H~I) n(e)}}  \\
 & & \propto \mathrm{(C/H)(Fe/H)^{-0.4}g^{-0.5}}.
\end{eqnarray}
with the approximation of n(e) and n(H)/n(e) in the line-forming region being
(Kj{\ae}rgaard et al., 1982)
\begin{eqnarray}
\mathrm{n(e)} & \pra & \mathrm{(Fe/H)^{0.4} g^{0.5}},\ \mathrm{and}\\
\mathrm{n(H)/n(e)} & \pra & \mathrm{(Fe/H)^{-1}}.
\end{eqnarray}

The approximate absorption coefficient of SiO is de\-fined as
\begin{eqnarray}
\lefteqn{\mathrm{{\cal{R}}(SiO) \equiv
\frac{\kappa(SiO)}{\kappa(H^-_{\mathrm{ff}})}}}  \\
 & & \propto \mathrm{\frac{n(SiO)}{n(H^-)}}  \\
 & & = \mathrm{\frac{n(Si~I) n(O~I) K(H^-)}{K(SiO) n(H~I) n(e)}}.
\end{eqnarray}
For oxygen-rich giants, n(O~I) may be approximated by
\[\mathrm{n(O~I) \approx n(O)-n(CO) \approx
n(O)-n(C)}.\]
The equilibrium of silicon is dominated by SiO and
Si~I (Si~II is present in the deeper layers as well). For example,
in the model with \teff =  4050 K, $\log$ g = 1.00, z = 0.00, M =
1.5 \Msun, \vt\ = 2.0 \km, the proportion of the number densities
are given in Table \ref{tablesio}. Since the ratio $\mathrm{n(SiO)/n(Si~I)}$
changes rapidly in the outer layers of the photosphere, a rough approximation to
$\mathrm{n(Si~I)}$, valid in the region where $-2 \le \log(\tau_0) \le -1$, is
then \\
\[\mathrm{n(Si~I) \approx (Si/H) n(H~I)},\]
so that
\begin{eqnarray}
\lefteqn{\mathrm{{\cal{R}}(SiO)}  \propto  \mathrm{(Si/H)(O/H-C/H) \frac{n(H~I)}{n(e)}
\frac{K(H^-)}{K(SiO)}}}  \\
 & & \propto \mathrm{(Si/H)(O/H-C/H)(Fe/H)^{-1}}.
\end{eqnarray}

\begin{table}[h]
\caption{\label{tablesio} Proportion of number densities for silicon for the
model with parameters \teff\ = 4050 K and $\log$ g = 1.00.}
\vspace{1ex}
\begin{center}
\begin{tabular}{r|rrr}\hline \hline
\rule[-3mm]{0mm}{8mm}$\log$($\tau_0$) & SiO & Si~I & Si~II \\
\hline
\rule[-0mm]{0mm}{5mm} $-5$ & 83 & 17 & 0 \\
$-4$ & 54 & 46 & 0 \\
$-3$ & 26 & 74 & 0 \\
$-2$ & 9 & 90 & 1 \\
$-1$ & 3 & 93 & 4 \\
\rule[-3mm]{0mm}{3mm}0 & 0 & 35 & 65 \\
\hline
\end{tabular}
\end{center}
\end{table}

Using the same notations, ${\cal{R}}$(OH) is written as
\begin{eqnarray}
\lefteqn{\mathrm{{\cal{R}}(OH) \equiv
\frac{\kappa(OH)}{\kappa(H^-_{\mathrm{ff}})}}}  \\
 & &  \propto \mathrm{\frac{n(OH)}{n(H^-)}}  \\
 & &  = \mathrm{\frac{n(O~I) n(H~I) K(H^-)}{K(OH) n(H~I) n(e)}}  \\
 & & \propto \mathrm{(O/H-C/H) \frac{n(H~I)}{n(e)} \frac{K(H^-)}{K(OH)}}  \\
 & & \propto \mathrm{(O/H-C/H) (Fe/H)^{-1}}.
\end{eqnarray}

Also for H$_2$O we obtain
\begin{eqnarray}
\lefteqn{\mathrm{{\cal{R}}(H_2O) \equiv
\frac{\kappa(H_2O)}{\kappa(H^-_{\mathrm{ff}})}}}   \\
 & & \propto \mathrm{\frac{n(H_2O)}{n(H^-)}}  \\
 & & = \mathrm{\frac{n(H_2) n(O~I) K(H^-)}{K(H_2O) n(H~I) n(e)}}  \\
 & & = \mathrm{\frac{n(H~I) n(H~I) n(O~I) K(H^-)}{K(H_2) K(H_2O) n(H~I) n(e)}}
  \\
 & & \propto \mathrm{(O/H - C/H) \frac{n^2(H~I)}{n(e)}}  \\
 & & = \mathrm{(O/H - C/H) (Fe/H)^{-1.6} g^{0.5}}.
\end{eqnarray}

The approximate absorption coefficient of NH is written as
\begin{eqnarray}
\lefteqn{\mathrm{{\cal{R}}(NH) \equiv
\frac{\kappa(NH)}{\kappa(H^-_{\mathrm{ff}})}}} \\
& & \propto \mathrm{\frac{n(NH)}{n(H^-)}}\\
& &  = \mathrm{\frac{n(N~I) n(H~I) K(H^-)}{K(NH) n(H~I) n(e)}}.
\end{eqnarray}
The equilibrium of the nitrogen species is dominated by
the N$_2$ formation. So,
\[\mathrm{n(N~I)^2 = n(N_2) K(N_2)},\]
with
\[\mathrm{n(N_2) \approx 0.5 (N/H) n(H~I)},\]
resulting in \[\mathrm{n(N~I) \pra [(N/H) n(H~I) K(N_2)]^{1/2}}.\]
Consequently,
\begin{eqnarray}
\lefteqn{\mathrm{{\cal{R}}(NH)}  \propto  \mathrm{(N/H)^{0.5} n(H~I)^{0.5} n(e)^{-1}}}
\\
 & & \propto \mathrm{(N/H)^{0.5} \frac{n(H~I)^{0.5}}{n(e)^{0.5}} n(e)^{-0.5}} \\
 & & \propto \mathrm{(N/H)^{0.5} (Fe/H)^{-0.5} (Fe/H)^{-0.2} g^{-0.25}} \\
 & & = \mathrm{(N/H)^{0.5} (Fe/H)^{-0.7}g^{-0.25}}.
\end{eqnarray}

\begin{acknowledgements}LD acknowledges support from the Science Foundation of
Flanders. BP thanks S. Johansson and the staff of the atomic spectroscopy
division at Lund university for their hospitality during
part of this work. This research has made use of the SIMBAD database, operated
at CDS, Strasbourg, France.
\end{acknowledgements}

\end{document}